\documentclass[11pt,a4paper]{article}

\usepackage{dcolumn}
\usepackage{amsmath,amsfonts,amsthm,amssymb}
\usepackage[latin1]{inputenc}
\usepackage{graphicx}
\usepackage{indentfirst}
\usepackage[usenames]{color}
\usepackage[nottoc, notlof, notlot]{tocbibind}
\usepackage{latexsym}
\usepackage{qsymbols}
\usepackage{float}
\usepackage{rotating}

\newcommand{\be}{\begin{equation}}
\newcommand{\ee}{\end{equation}}
\newcommand{\ba}{\begin{align*}}
\newcommand{\ea}{\end{align*}}

\newcommand{\R}{\mathbb R}
\newcommand{\Prob}{\mathbb{P}}
\newcommand{\F}{\mathcal F}

\newcommand{\esp}{\mathbb{E}}

\newcommand{\tinf}{t \wedge \tau_R}

\newcommand{\nB}{|B|}
\newcommand{\nA}{|A|}

\newcommand{\sigmaUnder}{\underline{\sigma}}
\newcommand{\sigmaBar}{\overline{\sigma}}

\newcommand{\Td}{T-\delta}
\newcommand{\yBar}{\overline{y}}
\newcommand{\rhop}{\rho_{\perp}}

\newcommand{\etaUnder}{\underline{\eta}}
\newcommand{\etaBar}{\overline{\eta}}
\newcommand{\Xbar}{\overline{X}}
\newcommand{\Vbar}{\overline{V}}
\newcommand{\RD}{R_{\Delta}}

\newtheorem{theor}{Theorem}[section]
\newtheorem{lem}{Lemma}[section]
\newtheorem{prop}{Proposition}[section]
\newtheorem{rem}{Remark}[section]
\newtheorem{cor}{Corollary}

\numberwithin{equation}{section}

\title{Bounds on Stock Price probability distributions in Local-Stochastic Volatility models}
\author{
V. Bally  
\footnote{Universit\'e Paris-Est, Laboratoire d'Analyse et Math\'ematiques Appliqu\'ees UMR 8050 CNRS, 5 bd Descartes, 77454 Marne-la-Vall\'ee Cedex 2, France. \texttt{Vlad.Bally@univ-mlv.fr}}
\and S. De Marco
\footnote{Universit\'e Paris-Est, Laboratoire d'Analyse et Math\'ematiques Appliqu\'ees
UMR 8050 CNRS and Scuola Normale Superiore, Piazza dei Cavalieri 7, 56126 Pisa, Italy. \texttt{s.demarco@sns.it}}
}
\date{June 14, 2010}

\begin{document}

\maketitle

\abstract{
We show that in a large class of stochastic volatility models with additional skew-functions (local-stochastic volatility models) the tails of the cumulative distribution of the log-returns behave as
$\exp(-c |y|)$, where $c$ is a positive constant depending on time and on model parameters.
We obtain this estimate proving a stronger result: using some estimates for the probability that It\^o processes remain around a deterministic curve from \cite{BalFern}, we lower bound the probability that the couple $(X,V)$ remains around a two-dimensional curve up to a given maturity, $X$ being the log-return process and $V$ its instantaneous variance.
Then we find the optimal curve leading to the bounds on the terminal cdf.
The method we rely on does not require inversion of characteristic functions but works for general coefficients of the underlying SDE (in particular, no affine structure is needed).
Even though the involved constants are less sharp than the ones derived for stochastic volatility models with a particular structure (\cite{AP,KellRes,Refined}),  our lower bounds entail moment explosion, thus
implying that Black-Scholes implied volatility always displays wings in the considered class of models.
In a second part of this paper, using Malliavin calculus techniques,
we show that an analogous estimate holds for the density of the
log-returns as well.

\vspace{2mm}

\vspace{2mm}

\vspace{2mm}
}

\section{Introduction}

\noindent
We consider the following class of diffusions:
\be \label{e:class} 
\begin{aligned}
dX_t &= -\frac{1}{2} \eta(t, X_t)^2 f(V_t)^2 dt
        + \eta(t,X_t) f(V_t) dW^1_t
\\
\\
d V_t &= \beta(t, V_t) dt + \sigma(t, V_t) dW^2_t,
\end{aligned}
\ee
where $W^1$ and $W^2$ are two correlated Brownian motions
on some filtered probability space $(\Omega, \F,$ $(\F_t)_{t \ge 0}, \Prob)$.
The function $f$ usually being positive, the couple $(X,f(V))$ lives in $\R \times \R^+$:
when $X$ models the logarithm of the forward price of an asset and $V$ its instantaneous variance,
Eq. (\ref{e:class}) defines a so-called local-stochastic volatility model (LSV).
The function $\eta$ is the local volatility (or skew) function;
the autonomous process $V$ is the stochastic volatility.
Local-stochastic volatility models embed stochastic volatility models (when $\eta \equiv 1$) and have been intensively studied by the financial community in these last years, in particular when the appearance of derivatives whose value depends on the dynamics of the implied volatility demanded the introduction of more elaborate models.
Some authors have focused on the problem of how to design an efficient calibration strategy of such a model to the market smile, as Lipton \cite{Lipt} or Henry-Labord\`ere \cite{PH1,PH2}, others have given a particular attention to the small-time asymptotics of implied volatility, as Forde \& Jacquier in \cite{JacqLSV}.
Here we do not focus directly on the problem of the calibration of (\ref{e:class}) to market data but we rather take up on the issue of giving asymptotic estimates of the cumulative distribution and (when existing) of the density of the log forward price $X$.
This problem is (clearly) related to the estimation of option prices and the calibration of the model, in a precise sense that is to be cleared hereafter.

\subsection{Related work}

\noindent
Let us precise the context and motivation of our study.
Setting $F_t = F_0 e^{X_t}$, $F_0 >0$, then $F = (F_t; t \ge 0)$ satisfies
\[
F_t = \int_0^t F_s \eta(s,X_s) f(V_s) dW^1_s,
\]
hence $F$ is a positive It\^o local martingale.
Then, a simple application of Fatou's Lemma allows to show that $F$ is actually an \emph{integrable} supermartingale.
Nevertheless, is is well known that within the class of models (\ref{e:class}), some complications may arise: $F$ may fail to be a true martingale
(cf. Sin \cite{Sin}, Jourdain \cite{Jourd}, Lions \& Musiela \cite{LionsMus}) and the moments of $F_t$ of order $p>1$ may become infinite.
The latter phenomenon has been attentively studied by several authors (cf. again \cite{LionsMus} or
Andersen \& Piterbarg \cite{AP} and Keller-Ressel \cite{KellRes}): if on the one
hand the lack of moment stability calls for additional care (if, for example, one wants to manipulate variances), on the other hand the explosion of moments plays a crucial role in
the investigation of the shape of the implied volatility surface.
Let us recall the basics results concerning implied volatility and moment explosion briefly.
The model-implied volatility of a European call option with time to maturity $T$ and
strike $F_0 e^k$ is the unique non-negative solution $\sigma(T,k)$
to the equation
\be \label{e:implVolDef}
\esp[ ( F_0 e^{X_T} - F_0 e^k)^+ ]  = C_{BS} (k, T, \sigma(T,k))
\ee
where $C_{BS} (k, T, \sigma)$ is the price of a Black-Scholes call option of strike
$F_0 e^k$, maturity $T$ and volatility $\sigma$.
Market-implied volatility $\sigma_{market}(T,k)$ is obtained in the same way, replacing the lhs in
(\ref{e:implVolDef}) with the market option prices for different values of $k$ and $T$.
Moreover, let the critical exponents $p^*_T(X)$ and $q^*_T(X)$ of $e^{X_T}$ be given by
\[
p^*_T(X) = \sup \{p \ge 1 : \esp [e^{p X_T}] < \infty \},
\hspace{8mm} q^*_T(X) = \sup \{q \ge 0  : \esp [e^{-q X_T}] < \infty \}.
\]
Lee's moment formula \cite{Lee} relies the critical exponents and the asymptotic slopes of implied
variance (the square of implied volatility) in the following way:
\be \label{e:momForm}
\limsup_{k \to \infty} \frac{T \sigma(T,k)^2}{k} = \varphi(p^*_T(X)-1),
\hspace{8mm}
\limsup_{k \to -\infty} \frac{T \sigma(T,k)^2}{k} = \varphi(q^*_T(X)),
\ee
where $\varphi(x) = 2 - 4 (\sqrt{ x^2 + x} - x ), \varphi(\infty) = 0$.
The relations in (\ref{e:momForm}) are proved by Lee in a complete model-independent framework.
We mention here that Benaim and Friz \cite{BenFr} sharpened Lee's result: under some technical conditions, they relate the left hand sides in (\ref{e:momForm}) directly to the tail asymptotics of the cumulative distribution function
of $X_T$, giving sufficient conditions for the $\limsup$ to be a true limit.
As pointed out by Lee, Eq. (\ref{e:momForm}) is useful for \emph{model selection} purposes: since the market-implied variance smiles usually display ``wings'' (i.e. $k \to \sigma_{market}(T,k)^2$ has left and right asymptotes), so has to do the model-implied volatility, hence the exponential moments of the underlying process $X$ \emph{must} explode (otherwise $p_T^*(X)=q^*_T(X)=\infty$ and (\ref{e:momForm}) tells that the implied volatility flattens for large values of $|k|$).
Pushing things further, the moment formula can help \emph{model calibration}: 
the values of the slopes at the left hand sides of (\ref{e:momForm}) can be observed on market data for values of $|k|$ large enough; if on the other hand the critical exponents are known functions of the model parameters, the use of (\ref{e:momForm}) can provide reasonable initial guesses of parameters values.
This explains the major interest of some authors in the explicit computation of critical exponents, as in \cite{AP} or \cite{KellRes} for some classes of stochastic volatility models: as a result, the critical exponents are not available in closed form but can be straightforwardly obtained solving (numerically) a simple equation.

Of course moment explosion is linked to the tail behaviour of the distribution of $X_T$.
More precisely, if the law of $X_T$ admits a density and this density behaves as $e^{-c |y|}$ for $|y|\to\infty$ for some constant $c>0$, then positive and negative exponential moments of $X_t$ of order $p$ will explode for $p \ge c$.
Dragulescu \& Yakovenko \cite{YakDrag} first showed that the density of
the log-price does behave as $e^{-c |y|}$ in the Heston stochastic volatility model \cite{Hest} (which is obtained from (\ref{e:class}) setting $\eta(t,x)=1$, $f(v) = \sqrt{v}$, $\beta(t,v) = k(\theta-v)$, $\sigma(t,v) = \sigma$), exploiting the analytical computations that can be carried for the characteristic function of $X_T$.
The work of \cite{YakDrag} on the Heston model has been extended and sharpened with the addition of higher-order terms to the leading $e^{-c |y|}$,
at first by Gulisashvili and Stein \cite{GulSt} in the case of zero correlation and subsequently by
Friz et al. \cite{Refined}.
In \cite{Refined}, relying on affine principles applicable to the Heston model, the authors find sharp aymptotic estimates for the forward price density
and use them to obtain some fine corrections to Gatheral's SVI parametric model of implied variance.

\subsection{Assumptions and results}

\noindent
The main aim of this work is to show that the cumulative distribution of the log forward price and, when existing, its density, behave as
$\exp(-c |y|)$ for large $y$ in the following class of LSV models:
\begin{align} \label{e:ourClass1}
dX_t &= -\frac{1}{2} \eta(t, X_t)^2 V_t dt
        + \eta(t,X_t) \sqrt{V_t} dW^X_t 
\\ \notag
\\ \label{e:ourClass2}
d V_t &= \beta(t, V_t) dt + \sigma(t, V_t) \sqrt{V_t} dW^V_t, 
\end{align}
obtained from (\ref{e:class}) setting $f(v)=\sqrt{v}$.
We consider finite time horizon $T>0$, correlated Brownian motions
$d\langle W^X, W^V \rangle_t = \rho dt$ and deterministic initial conditions $X_0= 0$ and $V_0 > 0$.
Eq. (\ref{e:ourClass2}) for the variance process is given on the domain $\R_+ = [0,\infty)$,
i.e. a process $V$ satisfying (\ref{e:ourClass2}) is such that $\Prob(V_t \in [0,\infty), t  \in [0,T]) = 1$.
This class contains the Heston model and the ``universal volatility model'' (without the jump part) considered in \cite{Lipt}, but is much wider since it
allows for general coefficients $\beta, \sigma$ in the SDE of the variance.
While on the one hand we consider reasonable Lipschitz, boundedness and ellipticity conditions on the coefficients $\eta$ and $\sigma$ (but we allow the drift $\beta$ to be any measurable function with sub-linear growth), on the other hand we emphasize that the square-root factors in (\ref{e:ourClass1})-(\ref{e:ourClass2})  makes it impossible to apply classical Malliavin calculus techniques, or  any other method requiring diffusions with globally Lipschitz coefficients, to estimate the density of the solution.
We also remark that we do not have to deal here with the (possibly intricate) discussion on the existence and/or uniqueness of solutions to (\ref{e:ourClass1})-(\ref{e:ourClass2}): our results hold for \emph{any} couple of processes $(X,V) = (X_t, V_t; t \in [0,T])$ satisfying (\ref{e:ourClass1})-(\ref{e:ourClass2}) .
The situation where the diffusion coefficient of Eq. (\ref{e:ourClass2}) is replaced by $\sigma(t,v) v^p$ for a $p>0$ (thus embedding the class of models considered by Andersen \& Piterbarg in
\cite{AP}) will be the subject of future work.

Our main tool to estimate the terminal (at time $T$) cumulative distribution of $X$
is an estimation involving the whole trajectory of the couple $(X,V)$ up to time $T$.
In \cite{BalFern} Bally, Fernandez \& Meda provide estimates for the probability that an It\^o process remains in a tube of given radius around a given deterministic curve, asking some conditions
of local Lipschiz-continuity, local boundedness and local ellipticity on
the coefficients of the SDE.
As a result, the probability of staying in the tube is lower bounded by an integral functional of the curve itself, of the deterministic radius and the coefficients of the SDE.
In the framework of the model (\ref{e:ourClass1})-(\ref{e:ourClass2}), we are able to 
cast this functional in a simple form and then to
optimize over the possible curves and radii, obtaining a lower bound
which is in the good asymptotic range.
To state our main result, let us introduce the following objects : for $y \in \R$, let $\yBar$ and the one-dimensional curves $\tilde{x}_t, \tilde{v}_t, \tilde{R}_t$, $t \in [0,T]$, be given by
\be \label{e:curves}
\yBar = |y| + V_0; \hspace{10mm} \phi(t) = \frac{\sinh(t/2)}{\sinh(T/2)};
\ee
\[
\tilde{v}_t =
V_0 \Bigl( \sqrt{ \frac{\yBar}{V_0} } \phi(t) - e^{-T/2}\phi(t) + e^{-t/2}\Bigr)
^2; \hspace{3mm}
\tilde{x}_t = \text{sign}(y) ( \tilde{v}_t - V_0 ); \hspace{3mm}
\tilde{R}_t = \frac{1}{2} \sqrt{ (V_0\wedge1) \tilde{v}_t }
\]
where $\text{sign}(x) = 1$ if $x \ge 0$ and $\text{sign}(x) = -1$ if $x <0$.
Our main result is the following estimate:
\be \label{e:mainEstim}
\Prob\bigl( | (X_t, V_t) - (\tilde{x}_t, \tilde{v}_t)| \le \tilde{R}_t,
0 \le t \le T\bigr) \ge
\exp \Bigl( - c_T \psi(\rhop) \times |y| \Bigr)
\ee
for $|y|$ large enough, where $\psi$ is an explicit function (cf. (\ref{e:psi}) in section 2 for the precise expression) and $c_T$ is a strictly positive constant depending on the model parameters and explicitly on $T$ but not on $y$ nor on the correlation parameter $\rho$.
The curves $\tilde{x}_{\cdot}, \tilde{v}_{\cdot}, \tilde{R}_{\cdot}$ in (\ref{e:curves}) are the product of the optimization procedure we set up
and they appear as the solution to some Euler-Lagrange equations (given in section \ref{s:Lagr}).
We remark that the curve $\tilde{x}_{\cdot}$ ends up at $\tilde{x}_T = y$ while the terminal radius $\tilde{R}_T$ is proportional to $\sqrt{|y|}$: hence, dropping constants and simply writing
$
\Prob(|X_T - y| \le \sqrt{|y|}) \ge \Prob( | (X_T, V_T) - (\tilde{x}_T, \tilde{v}_T)| \le \tilde{R}_T)
$
and using (\ref{e:estimTubeExpl}), we obtain the desired lower bound for the terminal distribution
(this argument is made rigorous in Corollary \ref{c:cdf} in section \ref{s:implVol}).
Notice that the fact that the ``tube'' estimate (\ref{e:mainEstim}) is given for the couple $(X,V)$ is crucial in our framework: indeed, in order to estimate the behaviour of $X_T$ we need to have a control on the variance $V_t$ for all $t \in [0,T]$.
 
In a second part of this work, we go some further and, under some
additional hypotheses on the coefficients of (\ref{e:ourClass1})-(\ref{e:ourClass2}), we discuss the existence of a (possibly non continuous) density for the law of $X_T$
and show that the exponential estimate holds for this density as well.
This last step requires to work out some ``small balls'' estimates (cf. Proposition \ref{p:smallBalls} in section \ref{s:density}) and to employ the integration by parts formula of Malliavin calculus.
More precisely, we rely on a decomposition of $X_T$ as a Gaussian term plus a perturbation, following the idea of Bally \& Caramellino in \cite{BalCar}: we obtain the desired lower bound performing a somehow delicate operation of balance
between the leading Gaussian term and the perturbation, involving regularization of the non-Lipschitz coefficients in (\ref{e:ourClass1})-(\ref{e:ourClass2}) and some precise estimates of Sobolev norms of a diffusion from \cite{SDM}.
Our final estimates on the density $p_{X_T}$ of $X_T$ reads
\[
p_{X_T}(y) \ge
\frac1{M_T}
\exp \bigl( -e_{T} \psi(\rhop) |y| \bigr) 
\]
for $|y|>M_T$, where $M_T$ and $e_T$ are constants depending on model parameters
and explicitly on $T$.
If the density is not continuous, the inequality holds almost surely (see Theorem \ref{t:density} for the precise statement).

The paper is organized as follows. 
In section \ref{s:main} we give our working hypotheses and a detailed presentation of the main results.
In particular, in \ref{s:tubes} we prove estimate (\ref{e:mainEstim}) and in \ref{s:implVol} we state the corollary for the terminal cdf, the moment explosion and the implied volatility slopes.
In \ref{s:density} we give our results on the density of $X_T$.
Sections \ref{s:lowerProof} and \ref{s:proofs2} are devoted to the proofs of the results stated in section \ref{s:main}: Malliavin calculus only appears in section \ref{s:proofs2}, while all the tools employed in the previous are borrowed from stochastic calculus for It\^o processes.
Finally, the Appendix contains the proofs of the most technical material and a remainder of the main elements of Malliavin differential calculus.

\section{Main results} \label{s:main}

\noindent
In this section we give our working hypotheses and a detailed presentation of the main results.

Let us consider the class of models (\ref{e:ourClass1})-(\ref{e:ourClass2}).
For the ease of computations, we decorrelate the driving Brownian motions in the usual way
and rewrite (\ref{e:ourClass1})-(\ref{e:ourClass2}) as
\begin{align} \label{e:LSV1}
dX_t &= -\frac{1}{2} \eta(t, X_t)^2 V_t dt
        + \eta(t,X_t) \sqrt{V_t} (\rho dW^1_t + \rhop dW^2_t), \ \ t \le T
\\ \notag
\\ \label{e:LSV2}
d V_t &= \beta(t, V_t) dt + \sigma(t, V_t) \sqrt{V_t} dW^1_t, \ \ t \le T
\end{align}
were $(W^1_t, W^2_t; t \le T)$ is a two-dimensional standard Brownian motion.
We consider deterministic initial conditions $X_0 = 0$ and $V_0 > 0$, finite time horizon $T > 0$,
$\rho \in (-1,1)$ and we denote $\rhop := \sqrt{1 - \rho^2}$.
Eq. (\ref{e:LSV2}) for the variance process is given on the domain $\R_+ = [0,\infty)$,
i.e. a process $V$ satisfying (\ref{e:LSV2}) is such that $\Prob(V_t \in [0,\infty), t  \in [0,T]) = 1$.
We assume that the coefficients $\eta,\beta$ and $\sigma$ in (\ref{e:LSV1})-(\ref{e:LSV2})
satisfy the following conditions, for some $K > 1$:

\begin{itemize}
\item[(R)] \emph{(regularity)} $\eta : [0,T] \times \R \to \R$ and
$\sigma : [0,T] \times  [0, \infty) \to \R$ are Lipschitz-continuous functions, more precisely
\[
\begin{aligned}
&|\eta(s,x) - \eta(t,y)| \le K ( |x-y| + |s-t|)
\\
\\
&|\sigma(s,v) - \sigma(t,u)| \le K( |v-u| + |s-t|)
\\
\end{aligned}
\]
hold for every $(s,t,x,y) \in [0,T] \times [0,T] \times \R \times \R$, respectively
every $(s,t,v,u) \in [0,T] \times [0,T] \times [0,\infty) \times [0,\infty)$.

\item[(G)] \emph{(growth)} The mesurable function $\beta : [0,T] \times  [0, \infty) \to \R$
has sub-linear growth in $v$, more precisely
\[
| \beta(t,v) | \le K(1 + v)
\]
for every $(t,v) \in [0,T] \times [0,\infty)$.
Moreover, there exist constants $0 < \etaUnder < 1 < \etaBar$ and
$0 < \sigmaUnder < 1 < \sigmaBar$ such that
\[
\etaUnder \le \eta(t,x) \le \etaBar, \hspace{10mm}
\sigmaUnder \le \sigma(t,v) \le \sigmaBar
\]
hold for every $(t,x) \in [0,T] \times \R$, respectively $(t,v) \in [0,T] \times [0,\infty)$.
\end{itemize}

\begin{rem}
\emph{Despite of the boundedness condition on $\eta$, $\sigma$ given in (G), obviously none of the drift and diffusion coefficients in the system (\ref{e:LSV1})-(\ref{e:LSV2}) is bounded, because of the factors $V_t$ and $\sqrt{V_t}$.}
\end{rem}

\begin{rem}
\emph{In hypothesis (R), we could replace Lipschitz-continuity with respect to the couple $(t,x)$ (resp. $(t,v)$) with  Lipschitz-continuity with respect to the state variable $x$ (resp. $v$) and Holder-continuity of exponent $1/2$ with respect to time, and all the results of sections \ref{s:tubes}
and \ref{s:implVol} would still hold.}
\end{rem}

\begin{rem}
\emph{We are not interested here in discussing the existence and/or uniqueness 
of solutions to (\ref{e:LSV1})-(\ref{e:LSV2}) under conditions (R) and (G).
All the results we give in this subsection and in the following
(but not in subsection \ref{s:density}, where a new set of hypotheses is
considered) indeed
hold for \emph{any} couple of processes $(X,V) = (X_t, V_t;  t \in [0,T])$
which satisfy (\ref{e:LSV1})-(\ref{e:LSV2}).}
\end{rem}

\paragraph{Notation}
\emph{Sets and filtrations.}
All along the paper, $|\cdot|$ will denote the absolute value for real numbers as well as
the Euclidean norm for vectors, i.e. $|x| = \sqrt{\sum_i^n x_i^2}$ if $x \in \R^n$.
We denote $B_R(x)$ the open ball in $\R^n$ of center $x$ and radius $R$, $B_R(x) = \{y \in R^n : |y-x|<R \}$.
Moreover, we denote $(\F^i_t, t \ge 0)$ the completion of the filtration generated by
$W^i$, $i = 1,2$, and $\F_t = \F^1_t \vee \F^2_t$.
$\lambda_n$ denotes the Lebesgue measure on $\R^n$.
\\
\emph{Classes of functions, derivatives, norms.}
$C^1([0, T])$ denotes the class of real functions of $[0,T]$ which have uniformly continuous
derivative on $(0,T)$.
We will make use of the class $L(\mu,h)$ defined in \cite{BalFern}, section 2 : given a fixed time
horizon $T$, $\mu \ge 1$ and $h > 0$, $L(\mu,h)$ is the class of functions
$f : [0,T] \to R_+ = [0,\infty)$ such that for every $t,s \in [0,T]$ with $|t-s| < h$ one has
\[
f(t) \le \mu f(s).
\]
We denote $C^{0,k}([0,T] \times \R^n)$ (resp. $C^{0,k}_b([0,T] \times \R)$)
the class of continuous functions of $[0,T] \times \R^n$ which have continuous
(resp. bounded continuous) partial derivatives with respect to the second variable up to
order $k$.
Let $\Theta_k$ be the set of multi-indexes of length $k$ with components in
$\{1, \dots, n\}$, $\Theta_k = \{1, \dots, n\}^k$.
For $\alpha \in \Theta_k$ and $g \in C^{0,k}([0,T] \times \R^n)$ we denote $\partial_{\alpha} g = \frac{ \partial^k g}{\partial_{x^{\alpha_1}} \cdots \partial_{x^{\alpha_k}}}$.
We define the norms
\[
|g|_k = 1 \vee \sum_{j=0}^k \sum_{\alpha \in \Theta_k} \sup_{t \in [0,T]} \sup_{x \in \R^n}
|\partial_{\alpha} g(t,x)|.
\]
\emph{Constants}.
For a vector of parameters $\Lambda = (\lambda_1, \dots, \lambda_{\nu})$, we shall denote $C_{\Lambda}$
(resp. $C_{\Lambda}(t)$) a positive constant (resp. function of time) depending on the $\lambda_i$'s but
not on any of the other existing variables.
All constants of such a type may vary from line by line, but always depend only on $\Lambda$.

\subsection{Estimates around a Deterministic Curve} \label{s:tubes}
\noindent
We fix $T >0$.
We consider three one-dimensional curves $x, v, R$ of class $C^1([0,T])$ such that: $R_t > 0$
for any $t \in [0,T]$ and $x$ and $v$ have the same initial values as $X$ and $V$ in
(\ref{e:LSV1})-(\ref{e:LSV2}) (in particular, $x_0 = 0$).
We look for a lower bound on the probability that a process
$(X_t,V_t) = (X_t, V_t; t \le T)$ satisfying (\ref{e:LSV1})-(\ref{e:LSV2})
stays in the tube of radius $R_t$ around the deterministic curve $(x_t, v_t)$ up to
time $T$, that is a lower bound on the quantity
\be \label{e:probaTube}
\Prob( | (X_t, V_t) - (x_t, v_t)| \le R_t, 0 \le t \le T).
\ee
To lower bound (\ref{e:probaTube}) we employ the estimate provided in \cite{BalFern}, Theorem 1.
The main result of this section, Theorem \ref{t:lowerTube}, makes use of the
triplet of curves $\tilde{x}_t, \tilde{v}_t, \tilde{R}_t$ defined in (\ref{e:curves}).
As addressed in the Introduction, the choice of these particular curves relies on an optimization
problem: they indeed appear as the
solutions of some Euler-Lagrange equations (see section \ref{s:Lagr}).
Recall the $\psi$ from the Introduction:
\be \label{e:psi}
\psi(r) = \frac1{r^6} \Bigl(\ln \Bigl( \frac{1}{r}\Bigr) + 1 \Bigr), \ \ \ r > 0. 
\ee

\begin{theor} \label{t:lowerTube}
Assume conditions \emph{(R)} and \emph{(G)} and let $(X_t,V_t;0 \le t \le T)$ be two processes satisfying (\ref{e:LSV1})-(\ref{e:LSV2}).
Then
for every $y \in \R$ with $|y|$ large enough, precisely
\be \label{e:yCond}
|y| > V_0 \bigl( 1 + 2 \sinh(T/2) \bigr)^2,
\ee
and with the curves $\tilde{x}_{\cdot}, \tilde{v}_{\cdot}, \tilde{R}_{\cdot}$ defined in (\ref{e:curves})
one has
\be \label{e:estimTubeExpl}
\Prob\bigl( | (X_t, V_t) - (\tilde{x}_t, \tilde{v}_t)| \le \tilde{R}_t,
t \in [0,T]) \ge
\exp \bigl( - c_T \psi(\rhop) \times |y| \bigr).
\ee
The constant $c_T$ is given by
\be \label{e:timeConst}
c_T = c^*  \Bigl( \frac1T + 1 \Bigr) e^{c^* T^2	},
\ee
where $c^*$ is strictly positive constant depending on the model parameters
$V_0, K, \etaUnder, \sigmaUnder, \etaBar, \sigmaBar$ given in \emph{(R)} and \emph{(G)} but not on $y$ nor on the correlation parameter $\rho$.
\end{theor}

\begin{rem}
\emph{Let us discuss
the impact of the factor $\psi(\rhop)$ and of the maturity $T$ in the lower bound (\ref{e:estimTubeExpl})
a bit further.
It is known that the correlation effects moment explosion in stochastic volatility models, a negative correlation bringing - as intuitively clear - a dampening effect (cf. \cite{AP}, sections 3 and 4).
In a Heston model, obtained when the variance process in (\ref{e:LSV2}) has
constant parameters and mean-reverting drift, the upper critical moment of $e^{X_T}$ tends to infinity when $\rho \to -1$ and $X_T$ even becomes a bounded
random variable when $\rho = -1$, and the behaviour is the opposite when $\rho > 1$.
The factor $\psi(\rhop)=\psi(\sqrt{1-\rho^2})$ has the expected explosive
behaviour when $\rho \to -1$, but 
it symmetrically decrements the rhs of (\ref{e:estimTubeExpl}) for $\rho > 0$,
making the lower bound significant in particular for $\rho \in (-1,0)$.
The small time asymptotics $\frac1T$ of the constant $c_T$ is what expected for a diffusion; on the other hand, the large time dependence $e^{c^* T^2}$
makes the bound (\ref{e:estimTubeExpl}) not directly applicable to study the large-time asymptotics.}
\end{rem}

\subsection{Lower bounds for Cumulative Distribution Function and Moments} \label{s:implVol}

\noindent
Theorem \ref{t:lowerTube} leads, in particular, to lower bounds on
the tails of the complementary cumulative distribution function (complementary cdf in short) of $X_T$, i.e. $\Prob(|X_T| > \cdot)$.
Indeed, on the one hand we can simply lower bound the probability to
be in the tube at the final ``time-slice'', $\Prob( | (X_T, V_T) - (\tilde{x}_T, \tilde{v}_T)| \le \tilde{R}_T)$ with the probability to stay in the tube up to time $T$.
On the other hand, the final time radius $\tilde{R}_T$ in (\ref{e:curves}) is - roughly
speaking - proportional to $\sqrt{|y|}$.
Hence, when $y \to \infty$ (resp. $y \to -\infty$) the infimum (resp. the supremum) of the interval $[y - \tilde{R}_T, y + \tilde{R}_T]$ becomes large (resp. small) and this allows to obtain tail estimates for $\Prob(|X_T| > \cdot)$ that are
in the same asymptotic range as (\ref{e:estimTubeExpl}).
This observation is made rigorous in the proof of the following Corollary, which is indeed a direct consequence of Theorem \ref{t:lowerTube}.

\begin{cor} \label{c:cdf}
Under the assumptions of Theorem \ref{t:lowerTube},
for any $y > 0$ satisfying (\ref{e:yCond}) and
\be \label{e:yCond2}
y > 2 (V_0 \vee 1)^2 (1 + V_0),
\ee
one has
\be \label{e:cdf}
\Prob( X_T > y ) \wedge \Prob( X_T < - y ) \ge
\exp \bigl( -c_T \psi(\rhop) \times y \bigr)
\ee
where $c_T$ is the constant given in (\ref{e:timeConst}).
In particular, the critical exponents are finite:
\be \label{e:momExpl}
p^*_T(X) \vee q^*_T(X) \le c_T \psi(\rhop),
\ee
hence the implied volatility diplays left and right wings, i.e.
\be \label{e:implVolAs}
\begin{aligned}
&\limsup_{k \to \infty} \frac{T \sigma(T,k)^2}{k} \ge \varphi(c_T \psi(\rhop)-1) > 0,
\\
&\limsup_{k \to -\infty} \frac{T \sigma(T,k)^2}{k} \ge \varphi(c_T \psi(\rhop)) >0.
\end{aligned}
\ee
\end{cor}

\begin{rem}
\emph{As addressed in the Introduction, $e^{X_T}$ is integrable for every $T>0$.
A simple application of Markov's inequality shows that, for every $y>0$, $\Prob(X_T>y)
= \Prob(e^{X_T}>e^y) \le e^{-y} \esp[e^{X_T}]$.
This is not in contradiction with (\ref{e:cdf}), because on the one hand
$\psi(\rhop)$ is greater than or equal to one for any value of
$\rhop \in (0,1]$ (cf. (\ref{e:psi})), and on the other the constant $c^*$ in (\ref{e:timeConst})
is greater than $1$, hence $c_T > 1$ for every $T>0$, too.
}
\end{rem}

\begin{rem}
\emph{In this paper we are mainly interested in the law of $X_T$.
Estimate (\ref{e:estimTubeExpl}) can of course be applied to derive the
analogous lower bound for the \emph{joint} law of $X_T$ and $V_T$.
See Proposition \ref{p:smallBalls} in the next section for a refined statement in this direction.}
\end{rem}

\subsection{Lower bounds for the density} \label{s:density}

\noindent
We consider now some stronger regularity conditions on the coefficients of
(\ref{e:LSV1})-(\ref{e:LSV2}):

\begin{itemize}
\item[(R')] (\emph{regularity'}) (R) and (G) hold and
$\eta \in C^{0,2}_b([0,T] \times \R)$, $\sigma \in C^{0,2}_b([0,T] \times [0,\infty))$,
$\beta \in C^{0,2}_b([0,T] \times [0,\infty)) \cap Lip([0,T] \times [0,\infty))$
with $|\eta|_2 \vee |\sigma|_2 \le K$.
\end{itemize}

\begin{rem}
\emph{Under condition (R'), the system (\ref{e:LSV1})-(\ref{e:LSV2})
admits a unique strong solution.
Indeed, the existence of a weak solution $(X,V)$ that satisfies 
(\ref{e:preliminary}) follows from the continuity and sub-linearity
of the coefficients.
Then, pathwise uniqueness holds for (\ref{e:LSV2}) after a theorem of uniqueness
of Yamada and Watanabe (cf. \cite{KS}, Prop. 5.2.13) and
weak existence and pathwise uniqueness together imply strong existence
(\cite{KS}, Cor. 5.3.23).
Given the unique solution to (\ref{e:LSV2}), standard arguments allow to
prove pathwise uniqueness for (\ref{e:LSV1}).}
\end{rem}

We will give a lower bound for the density of the law of $X_T$ under hypothesis (R').
Notice first that the law of $X_T$ is absolutely continuous with respect to the Lebesgue
measure $\lambda_1$ on $\R$.
This fact may be proven in (at least) two ways.
First, we may look to the law of $X_T$ conditional to $(W^1_t, t \le T)$.
Then $X_T$ appears as a functional of the independent Brownian motion $(W^2_t, t \le T)$
and, using the Bouleau-Hirsch criterium (cf. \cite{Nual06}), we obtain a density
$p_{X_T}(W^1, x)$ for the conditional law.
Then, the law of $X_T$ has the density $\esp[ p_{X_T}(W^1,x)] = p_{X_T}(x)$.
A second way would be to use the results in \cite{SDM} (Theorem 2.2)
telling	 that the \emph{couple} $(X_T, V_T)$ admits a density $p_T(x,v)$ on $\R \times (0,\infty)$ (meaning that the law of $(X_T, V_T)$ restricted to $\R \times (0,\infty)$ has the density $p_T(x,v)$).
This immediately yields the
existence of a density $p_{X_T}(x)$ for the marginal law of $X_T$.
Nevertheless, we remark that none of the above approaches guarantee that the density 
of $X_T$ is continuous.

Before giving an estimate of the density of $X_T$ itself, we need to work
out some estimates for the probability that $X_T$ stays in a ball of ``small'' radius.

\begin{prop}[Lower bounds for balls of small radius] \label{p:smallBalls}
Let $R^{(j)}(y)$ be given by
\[
R^{(j)}(y) = \bigl( \sqrt{|y|} \bigr)^{1-j}, \ \ \ j \in \mathbb{N}
\]
(so that $R^{(0)}(y) = \sqrt{|y|}$, $R^{(1)}(y) = 1$, $R^{(2)}(y) = \frac1{\sqrt{|y|}}$, ...).
Assume (R') and let $(X_t, V_t; 0 \le t \le T)$ be the unique strong solution to
(\ref{e:LSV1})-(\ref{e:LSV2}).
Then,
for any $y$ satisfying (\ref{e:yCond}) and $|y|> 16 \vee 2 (V_0 \vee 1)^2 (1 + V_0)$,
\be \label{e:smallBalls}
\Prob\bigl( | (X_T,V_T) - (y,|y|+V_0) | \le R^{(j)}(y) \bigr) \ge
\exp \bigl( - (j+1) d_T \psi(\rhop) \times |y| \bigr).
\ee
The constant $d_T$ is given by $d_T = 2 c^* \Bigl(\frac1{T^2} + 1\Bigr) e^{(c^* +1)T^2}$,
$c^*$ being the constant in (\ref{e:timeConst}).
\end{prop}

\begin{rem}
\emph{By taking $j$ large enough, the radius $R^{(j)}(y)$
can be made arbitrarily small.
Then we would like to make use of (\ref{e:smallBalls}) to obtain a lower bound for the density of $X_T$ computed at $y$, but we cannot pass to the limit with $j$ in (\ref{e:smallBalls}) because the rhs tends to zero as $j \to \infty$.
Nevertheless, we can obtain a lower bound for the density using (\ref{e:smallBalls}) for finite $j$ and the integration by parts formula of Malliavin Calculus.
This is what we actually do in order to prove the next theorem.}
\end{rem}

Here is the main result for this section.

\begin{theor} \label{t:density}
Assume (R') and let $(X_t, V_t; 0 \le t \le T)$ be the unique strong solution
to (\ref{e:LSV1})-(\ref{e:LSV2}).
Then, there exists a strictly positive constant $M_T$
depending on $T$ and on the model parameters
such that for $\lambda_1$-a.e. $y$ with $|y|>M_T$,
\be \label{e:density}
p_{X_T}(y) \ge
\frac1 M_T
\exp \bigl( -e_{T} \psi(\rhop) |y| \bigr) 
\ee
where $e_T =136 c^* \Bigl( \frac1{T^2} + 1 \Bigr) e^{(c^*+1)T}$.
The inequality (\ref{e:density}) is understood in the sense
\[
\int_{|y| > M_T} f(y) p_{X_T}(y) dy \ge
\frac1 M_T \int_{|y| > M_T} f(y) \exp \bigl( -e_{T} \psi(\rhop) |y| \bigr) dy
\]
for every $f \in C_b(\R)$.
\end{theor}

\begin{rem}
\emph{If the density $p_{X_T}(y)$ is continuous, then (\ref{e:density}) holds for every $y$
with $|y|>M_T$.}
\end{rem}

\noindent
We recall that for a Heston model with constant coefficients, the density
$p_{X_T}(y)$ is asymptotic to $\exp(-c |y|)$, cf. \cite{YakDrag, Refined}.

\section{Proof of results in \ref{s:tubes} and \ref{s:implVol}} \label{s:lowerProof}
 
\noindent
We start by giving a preliminary result that will be used in the proof of Theorem
\ref{t:lowerTube}.
We consider $x_{\cdot}, v_{\cdot}$ in $C^1([0,T])$ and $R,c,\lambda,\gamma,L : [0,T] \to \R_+$ satisfying
\be \label{e:conditionCurves}
\begin{aligned}
&x_0 = 0; \hspace{5mm} 
v_0 = V_0; \hspace{5mm} v'_t > 0;
\\
&x', v', R, c,\lambda,\gamma,L \in L(\mu, h)
\end{aligned}
\ee
and we define the stopping time
\[
\tau_R = \tau_R(X,V) = \inf\{t \le T : |(X_t,V_t)-(x_t,v_t)| > R_t \}.
\]
Moreover, we denote
\[
b(t,x,v) = \left(\begin{array}{c} -\frac{1}{2}\eta(t,x)^2 v \\ \beta(t,v) \end{array}\right);
\]
\[
\begin{array}{c c}
\sigma_1(t,x,v) = \left(\begin{array}{c} \rho \eta(t,x) \sqrt{v} \\
\sigma(t,v) \sqrt{v} \end{array}\right);
&
\sigma_2(t,x,v) = \left(\begin{array}{c} \rhop \eta(t,x) \sqrt{v} \\ 0 \end{array}\right)
\end{array}
\]
and consider the conditions:
\begin{align} \label{e:H1}
&| b (t,X_{\tinf},V_{\tinf}) | + \sum_{j = 1, 2} | \sigma_j(t,X_{\tinf},V_{\tinf}) | \le c_t;
\\ \label{e:H2}
&\lambda_t I_2 \le \sigma \sigma^* (t, X_{\tinf}, V_{\tinf}) \le \gamma_t I_2;
\\ \label{e:H3}
&\esp \Bigl[
\sum_{j = 1, 2} | \sigma_j( s, X_s, V_s ) - \sigma_j( t, X_t, V_t ) |^2
1_{ \{ \tau_R \ge s \} } 
\Bigr] \le L_t^2 (s-t)
\end{align}
which correspond to hypothesis \textbf{(H)} in \cite{BalFern}.
Then, according to Theorem 1 in \cite{BalFern}, the estimate 
\be \label{e:probaTubeEstim}
\Prob( | (X_t, V_t) - (x_t, v_t)| \le R_t, 0 \le t \le T) \ge
\exp \Bigl( - Q(\mu) \Bigl(1 + \int_0^T  F_{x,v,R} (t) dt \Bigr) \Bigr)
\ee
holds with the \emph{rate function}
\be \label{e:rateFunction}
F_{x,v,R} (t) =
\frac{1}{h}
+ \frac{( x'_t )^2 + ( v'_t )^2}{\lambda_t}
+ 2 (c_t^2 + L_t^2 ) \Bigl( \frac{1}{\lambda_t} + \frac{1}{R_t^2} \Bigr).
\ee
and the constant $Q(\mu)$ given by
\be \label{e:Q}
Q(\mu) = \frac{q_{\mu}}{\phi_{\lambda, \gamma}^2}
\ln \frac{q_{\mu}}{\phi_{\lambda, \gamma}},
\ee
where
\be \label{e:phi}
\phi_{\lambda, \gamma} = \inf_{t \le T} \frac{\lambda_t}{\gamma_t};
\hspace{5mm}
q_{\mu} = 8^{12} e^2 \mu^{73}.
\ee
(We actually denote $\phi_{\lambda, \gamma}$ the constant $\rho$ in \cite{BalFern}).
The following proposition is the starting point to prove Theorem \ref{t:lowerTube}.

\begin{prop} \label{p:tech}
Assume conditions \emph{(R)} and \emph{(G)}.
Let $x_t, v_t, R_t$ satisfy (\ref{e:conditionCurves}) and consider a process
$(X_t,V_t) = (X_t, V_t; 0\le t\le T)$
satisfying (\ref{e:LSV1})-(\ref{e:LSV2}).
Let moreover 
\be \label{e:radius1}
R_t \le R v_t,  \ \ \ t \in [0, T]
\ee
hold for a fixed $R \in (0, 1)$.
Then, setting $\Theta = (K, \etaBar, \sigmaBar, R, V_0)$, there exist strictly positive constants
$c = c_{\Theta}; L = L_{\Theta};
\gamma = \gamma_{\Theta}; \lambda = \lambda_{\Theta,\etaUnder, \sigmaUnder}$
such that for every $0\le t< s\le T$ the conditions (\ref{e:H1})-(\ref{e:H2})-(\ref{e:H3})
are fulfilled by the curves
\[
\begin{array}{l}
c_t = c v_t; \hspace{9mm} L_t^2 = L_T v_t;
\\
\\
\gamma_t = \gamma v_t; \hspace{9mm}
\lambda_t = \rhop^2 \lambda v_t.
\end{array}
\]
$L_T$ is given by $L_T = L e^{C_2 T^2}$, where $C_2$ is the constant appearing in Lemma (\ref{l:preliminary}).
The curves $c_t, L_t, \gamma_t, \lambda_t$ belong respectively to $L(\mu,h), L(\sqrt{\mu},h),
L(\mu,h)$, $L(\mu,h)$.
\end{prop}

\proof[Proof]{
In what follows we shall repeatedly apply the inequality $\sqrt{v} \le 1 + v$,
$v > 0$.
\\
(\ref{e:H1}): We notice that for every $t,x,v \in [0,T] \times \R \times [0,\infty)$,
\[
\begin{aligned}
| b (t,v,x) | + \sum_{j = 1, 2} &| \sigma_j(t,v,x) |
\\
&\le
\frac{1}{2} \etaBar^2 v + K(1+v) + (\rho + \rhop) \etaBar \sqrt{v} + \sigmaBar \sqrt{v}
\\
&\le
K + (\rho + \rhop) \etaBar + \sigmaBar +
\Bigl( \frac{1}{2} \etaBar^2 + K + (\rho + \rhop) \etaBar + \sigmaBar \Bigr) v
\\
&\le
c(1 + v)
\end{aligned}
\]
where the last holds with $c = \frac{1}{2} \etaBar^2 + K + 2 \etaBar + \sigmaBar$.
Then, employing the condition (\ref{e:radius1}) on the radius and the fact that
$v_t \ge V_0$ for any $t \in [0,T]$ by (\ref{e:conditionCurves}),
\be \label{e:C}
\begin{aligned}
| b (t,X_{\tinf},V_{\tinf}) | + \sum_{j = 1, 2} | &\sigma_j(t,X_{\tinf},V_{\tinf}) | 
\\
&\le c (1 + V_{\tinf}) \le c (1 + (v_t + R_t))
\\
&\le c (1+R) (1 + v_t) \le 2 c (1+R) \frac{1 \vee V_0}{V_0} v_t 
\end{aligned}
\ee
and the last inequality holds since $(1+v) \le 2 \frac{V_0 \vee 1}{V_0} v $ for any $v > V_0$.
\\
(\ref{e:H2}): Let $\sigma \sigma^*_{i,j}(t,x,v) = \sum_{k=1,2} \sigma^i_k (t,x,v) \sigma^j_k(t,x,v)$, $i,j = 1,2$.
The condition
\be \label{e:HI3}
\lambda_t I_2 \le \sigma \sigma^* (t, X_{\tinf}, V_{\tinf}) \le \gamma_t I_2
\ee
for the given $\lambda_t, \gamma_t$ will follow from the computation of the eigenvalues of
$\sigma \sigma^*$.
Denoting $\eta := \eta(t,x)$ and $\sigma := \sigma(t,v)$ for simplicity of notation, we have
\[
\sigma \sigma^* (t,x,v) =
\left(\begin{array}{c c}
\eta^2 v & \rho \eta \sigma v
\\ \rho \eta \sigma v & \sigma^2 v
\end{array}\right)
\]
hence the smallest, respectively the largest, eigenvalue satisfy
\be \label{e:lambdGamma}
\begin{aligned}
\overline{\lambda}_t(x,v) &=
\frac{1}{2} \Bigl(
\eta^2 v + \sigma^2 v -
\sqrt{ \bigl( \eta^2 v + \sigma^2 v \bigr)^2 - 4 \eta^2 \sigma^2 v^2 \rhop }
\Bigr)
\\
&\ge
\rhop^2 \frac{ \eta^2 \sigma^2 v^2 }
{\eta^2 v + \sigma^2 v}
\ge \rhop^2 \frac{\etaUnder^2 \sigmaUnder^2}{2(\etaBar^2 + \sigmaBar^2)}
v 
\\
\overline{\gamma}_t(x,v) &=
\frac{1}{2} \Bigl(
\eta^2 v + \sigma^2 v +
\sqrt{ \bigl( \eta^2 v + \sigma^2 v \bigr)^2 + 4 \eta^2 \sigma^2 v^2 \rhop }
\Bigr)
\\
&\le
\eta^2 v + \sigma^2 v
\le (\etaBar^2 + \sigmaBar^2) v
\end{aligned}
\ee
Proceeding as before we have
\[
\begin{aligned}
\overline{\lambda}_t(X_{\tinf},V_{\tinf}) &\ge
\rhop^2 \frac{2(\etaUnder^2 \sigmaUnder^2)}{\etaBar^2 + \sigmaBar^2}
\frac{1-R}{1+R} v_t =
\rhop^2 \lambda(\etaUnder, \sigmaUnder, \etaBar, \sigmaBar, R) v_t;
\\
\overline{\gamma}_t(X_{\tinf},V_{\tinf}) &\le
(\etaBar^2 + \sigmaBar^2) (1+R) v_t
=
\gamma(\etaBar, \sigmaBar, R)
v_t;
\end{aligned}
\]
Then (\ref{e:HI3}) follows with $\lambda_t$, $\gamma_t$ as in the statement of the proposition.
\\
(\ref{e:H3}):
Because of assumption (R), for every $s, t \in [0,T] \times [0,T]$, every $x, y \in \R \times \R$ and every
$v, u \in [0,\infty) \times [0, \infty)$ we have
\be \label{e:L1}
\begin{aligned}
| \eta(s,x) \sqrt{v} - \eta(t,y) \sqrt{u}| &\le
\sqrt{u} K ( |x-y| + |s-t| ) + \etaBar |\sqrt{v} - \sqrt{u}| 
\\
&\le
\sqrt{u} K ( |x-y| + |s-t| ) + \frac{\etaBar}{2 \min(\sqrt{v}, \sqrt{u})}|v - u|
\end{aligned}
\ee
and
\be \label{e:L2}
\begin{aligned}
| \sigma(s,v) \sqrt{v} - \sigma(t,u) \sqrt{u}| &\le
\sqrt{u} K ( |x-y| + |s-t| ) + \sigmaBar |\sqrt{v} - \sqrt{u}| 
\\
&\le
\sqrt{u} K ( |x-y| + |s-t| ) + \frac{\sigmaBar}{2 \min(\sqrt{v}, \sqrt{u})} |v - u|.
\end{aligned}
\ee
It follows, for every $t \le s \le T$,
\[
\begin{aligned}
\esp &\Bigl[ | \eta(s,X_s) \sqrt{V_s} - \eta(t,X_t) \sqrt{V_t}|^2 1_{ \{ \tau_R \ge s \} }\Bigr]
\\
&\le
4 K^2 \esp \bigl[ V_t \bigl( |X_s-X_t|^2 + (s-t)^2 \bigr) 1_{ \{ \tau_R \ge s \} }] +
\esp[ \frac{\etaBar^2}{2 \min(V_t,V_s)} |V_s - V_t|^2 1_{ \{ \tau_R \ge s \} } \bigr]
\\
&\le 
4 K^2 \Bigl( (1+R) v_t C_2 e^{C_2 T^2} (s-t) + T (s-t) \Bigr) + \frac{\etaBar^2}{2 (1-R) v_t} C_2 e^{C_2 T^2} (s-t)
\\
&\le 
C_2 \Bigl( 8 (1+R) K^2 \frac{V_0 \vee 1}{V_0} + \frac{\etaBar^2}{2 (1-R) V_0^2} \Bigr) e^{C_2 T^2} v_t (s-t)
\end{aligned}
\]
where $C_2$ is the constant considered in Lemma \ref{l:preliminary}.
Analogously,
\[
\begin{aligned}
\esp &\Bigl[ | \sigma(s,V_s) \sqrt{ V_s} - \sigma(t,V_t) \sqrt{V_t}|^2 1_{ \{ \tau_R \ge s \} }\Bigr]
\\
&\le
4 K^2 \Bigl( (1+R) v_t C_2 e^{C_2 T^2} (s-t) + T (s-t) \Bigr) + \frac{\sigmaBar^2}
{2(1-R) v_t} C_2 e^{C_2 T^2} (s-t)
\\
&\le 
C_2 \Bigl( 8 (1+R) K^2 \frac{V_0 \vee 1}{V_0} + \frac{\sigmaBar^2}{(1-R) V_0^2} \Bigr) e^{C_2 T^2} v_t (s-t).
\end{aligned}
\]
Estimate (\ref{e:H3}) then follows from the two previous inequalities and the
expression of $\sigma_1$, $\sigma_2$.
\\
The last statement on the curves $c_t, L_t, \gamma_t, \lambda_t$
follows from the fact that the function $a f^{p}$ belongs to $L(\mu^p,h)$ if $f$ belongs to
$L(\mu,h)$, $p > 0$ and $a$ is a positive constant. 
}
\endproof

Basically, what Theorem \ref{t:lowerTube} does is to compute the right hand side of
(\ref{e:probaTubeEstim}) on a particular curve satisfying conditions
(\ref{e:H1})-(\ref{e:H2})-(\ref{e:H3}), so that
(\ref{e:probaTubeEstim}) translates into the explicit lower bound (\ref{e:estimTubeExpl}).
The choice of the deterministic curve $(x_t, v_t)$ considered in Theorem \ref{t:lowerTube}
is of course motivated by the form of the rate function (\ref{e:rateFunction}).
More precisely, consider any $x_t, v_t, R_t$ that satisfy (\ref{e:conditionCurves}).
Then, by Proposition \ref{p:tech}, the estimate (\ref{e:probaTubeEstim}) holds with
\be \label{e:ourRateF}
F_{x,v,R} (t) =
\frac{1}{h}
+ \frac{( x'_t )^2 + ( v'_t )^2}{\rhop^2 \lambda v_t}
+ 2 (c^2 v_t^2 + L_T v_t ) \Bigl( \frac{1}{\rhop^2 \lambda v_t} + \frac{1}{R_t^2} \Bigr),
\ee
\be \label{e:ourRho}
\phi_{\lambda,\gamma} =
\inf_{t \le T} \frac{\lambda_t}{\gamma_t}
=\frac{\rhop^2 \lambda}{\gamma} \inf_{t \le T} \frac{v_t}{v_t}
= \frac{\rhop^2 \lambda}{\gamma}
\ee
and $Q(\mu)$ as given in (\ref{e:Q}).

\subsection{A Lagrangian minimization problem} \label{s:Lagr}

\noindent
We start from the simple observation that maximizing the lower bound in (\ref{e:probaTubeEstim}) is equivalent to minimizing the exponent
$Q(\mu)(1 + \int_0^T F_{x,v,R}(t)dt)$.
Due to the presence of the competing terms
$\frac1{ v_t } + \frac{1}{R_t^2}$
and $( x'_t )^2 + ( v'_t )^2$ in $F_{x,v,R}$, we make the choice
\be \label{e:radiusChoice}
R_t = \frac12 \sqrt{ V_0  v_t }
\ee
so that $R_t^2$ is proportional to $v_t$,
and consider curves $x_t, v_t$
such that $ |x'_t| = |v'_t|$, precisely
\be \label{e:shiftedCurv}
x_t = sign(y) ( v_t - V_0 ),  \ \ \ t \in [0,T].
\ee
Equations (\ref{e:radiusChoice}) and (\ref{e:shiftedCurv}) define
$R_t$ and $x_t$ given $v_t$, as happens for (\ref{e:curves}).
We remark that the radius in (\ref{e:radiusChoice}) satisfies the requirement $R_t \le \frac{1}{2} v_t$ of Proposition \ref{p:tech}.
Moreover, if the arrival point $x_T = y$ of the curve $x_t$ is given, the same will be for $v_t$.
We define the ``shifted'' arrival point $\yBar = v_T$ setting
\be \label{e:shiftedY}
\yBar = |y| + V_0.
\ee
After (\ref{e:radiusChoice})-(\ref{e:shiftedCurv}), the rate function
(\ref{e:ourRateF}) reduces to
\be \label{e:rateFmodif}
\overline{F}_{v} (t) =
\frac{1}{h}
+ \frac{2}{\rhop^2 \lambda v_t} ( v'_t )^2
+ 2 (c^2 v_t^2 + L_T v_t) \Bigl( \frac{1}{\rhop^2 \lambda} + 1\Bigr)
\frac{1}{v_t}
\ee
which is a function of the curve $v_t$ only.
Since we want to upper bound $\overline{F}_{v}$, we can get rid of all the constants and
just keep the explicit dependence with respect to the curve $v_t$: defining
\be \label{e:Gamma}
\Gamma_T = 1 \vee 2 \frac{ c^2 + L_T }{(V_0 \wedge 1)\rhop^2 \lambda},
\ee
we have 
\[
\overline{F}_{v} \le
\frac{1}{h}
+ \Gamma_T \Bigl(
\frac{( v'_t )^2}{v_t} + v_t 
\Bigr)
\]
and the constant $\Gamma$ carries the explicit dependence w.r.t the model
parameter $\rhop$.
The strategy we shall follow is to consider
\be \label{e:lagr}
\mathcal{L}(v_t, v'_t) = \frac{( v'_t )^2}{v_t}
+ v_t
\ee
and to look for the solution of the minimization problem
\be \label{e:lagrMin} 
\min_{v} \int_0^T \mathcal{L}(v_t, v'_t)
\ee
with the constraints
\be \label{e:constr}
v_0 = V_0; \hspace{10mm} v_T = \yBar.
\ee
The problem (\ref{e:lagrMin})-(\ref{e:constr}) is the classical minimization problem 
in Calculus of Variations for Lagrangian systems: a stationary point for the
integral functional in (\ref{e:lagrMin} ) is given by the solution of the Euler-Lagrange equation
\[
\frac{d}{dt} \frac{d \mathcal{L}}{dv'} (v_t, v'_t)
- \frac{d \mathcal{L}}{dv}(v_t, v'_t) = 0.
\]
under the constraints (\ref{e:constr}).
Some simple calculations yield the Euler-Lagrange equation associated to the Lagrangian
(\ref{e:lagr}): this equation reads
\be \label{e:explEulLagr}
\frac{ v''_t }{ v'_t }
= \frac{ v'_t }{ 2 v_t }
+ \frac{ v_t}{ 2 v'_t }.
\ee
A closer look to Eq. (\ref{e:explEulLagr}) reveals that it can be converted into a
linear second order ODE - hence explicitly solved - with the change of variables
\be \label{e:change}
u_t = \Bigl( \frac{v_t}{V_0} \Bigr)^{\frac12},
\ee
which indeed converts (\ref{e:explEulLagr}) into
\be \label{e:explEulLagr2}
u''_t - \frac14 u_t = 0,
\ee
now with the constraints
\be \label{e:constr2}
u_0 = 1; \hspace{10mm} u_T = \Bigl( \frac{\yBar}{V_0} \Bigr)^{\frac12}.
\ee
The explicit solution to (\ref{e:explEulLagr2})-(\ref{e:constr2}) is
easily found to be

\be \label{e:explSol}
u_t = \Bigl(\frac{\yBar}{V_0}\Bigr)^{\frac12} \frac{\sinh(t/2)}{\sinh(T/2)} 
- e^{- T/2} \frac{\sinh(t/2)}{\sinh(T/2)} + e^{-t/2}.
\ee

\bigskip
\noindent
The curve $\tilde{v}_t$ defined in (\ref{e:curves}) corresponds to the one given by
(\ref{e:change}) and (\ref{e:explSol}).
What Theorem \ref{t:lowerTube} does, then, is to pick up this particular curve,
to check for which values of $\mu, h$ and $y$ the curve $\tilde{v}'_t$ belongs to $L(\mu, h)$ and satisfies $\tilde{v}'_t > 0$ from (\ref{e:conditionCurves}),
hence to estimate the integral functional in (\ref{e:lagrMin}).


\proof[Proof of Theorem \ref{t:lowerTube}]{
\emph{Step 1.}
We show that $\tilde{v}'_t > 0$, $t \in [0,T]$ and $\tilde{v}' \in L(4, h)$
with $h = \Bigl(\frac{\yBar}{V_0}\Bigr)^{\frac12} \tanh(T/2)$, if y satisfies (\ref{e:yCond}).
Taking advantage of the notation introduced in (\ref{e:change}), we have
\[
\tilde{v}'_t = 2 V_0 u_t u'_t.
\]
and a simple calculation yields
\be \label{e:uDeriv}
u'_t = \Bigl(
\Bigl(\frac{\yBar}{V_0}\Bigr)^{\frac12} - e^{-T/2} \Bigr)
\frac{\cosh(t/2)}{2\sinh(T/2)} - \frac12 e^{-t/2}.
\ee
We remark that $u_t > 0$ for every $t \in [0,T]$ as soon as $y > X_0$, hence by (\ref{e:explEulLagr2})
$u''_t > 0$, too, and consequently $u'_t$ is an increasing function.
Using the expression of $u'_0$ given by (\ref{e:uDeriv}), it is easy to verify that (\ref{e:yCond}) implies $u'_t \ge u'_0 \ge \frac14 > 0$.
Now, we simply observe that $u \in L(2, ||u'||^{-1}_{\infty})$:
indeed, for every $s,t \in [0,T]$ such that $|s-t| < ||u'||^{-1}_{\infty}$,
\[
u_s \le u_t + ||u'||_{\infty} |s-t|
\le u_t + 1 \le 2 u_t
\]
where the last holds because $u_t \ge u_0 = 1$.
Analogously, $u' \in L(2, ||u||^{-1}_{\infty})$ because
\[
u'_s \le u'_t + ||u''||_{\infty} |s-t|
\le u'_t + (1-a)^2 ||u||_{\infty} |s-t|
\le 2 u'_t
\]
holds if $|s-t| < ||u||^{-1}_{\infty}$, employing in the last step the fact that
$u'_t \ge u'_0 \ge \frac14$.
Because $u_t$ and $u'_t$ are increasing, we have
\[
|| u ||_{\infty} = u_T = \Bigl( \frac{\yBar}{V_0} \Bigr)^{\frac12};
\hspace{5mm}
|| u' ||_{\infty} = u'_T
\le \Bigl(\frac{\yBar}{V_0}\Bigr)^{\frac12} \frac 1{2\tanh(T/2)}
\]
Observing that $\tanh(T/2) < 1$, we conclude that both $u$ and $u'$ belong to the class $L(2,h)$ with $h = \Bigl(\frac{V_0}{\yBar} \Bigr)^{\frac12} \tanh(T/2)$.
The fact that $\tilde{v}'_t \in L(4, h)$ for the same $h$ now follows from
the property $c f g \in L(\mu_f \times \mu_g, h_f \wedge h_g)$ if $f \in L(\mu_f, h_f)$,
$g \in L(\mu_g, h_g)$ and $c$ is a constant.

\emph{Step2.}
We estimate the integral functional at the right hand side of (\ref{e:probaTubeEstim}).
\\
By Proposition \ref{p:tech} and the computations at the beginning of the current section,
we know that the rate function $F_{\tilde{x},\tilde{v},\tilde{R}}$ is upper bounded
by $\overline{F}_{\tilde{v}}$ defined in (\ref{e:rateFmodif}), more precisely
$ F_{\tilde{x},\tilde{v},\tilde{R}} \le \frac{1}{h} + \Gamma_T \Bigl(
\frac{(\tilde{v}'_t )^2}{ \tilde{v}_t} + \tilde{v}_t \Bigr). $
Making once again use of $u$ defined in (\ref{e:change}), we have
$(\tilde{v}'_t)^2 = 4 V_0^2 u_t^2 (u'_t)^2$, hence
\[
\begin{aligned}
\int_0^T F_{\tilde{x},\tilde{v},\tilde{R}} (t) dt
&\le
\int_0^T \Bigl( 
\frac{1}{h} + \Gamma_T \Bigl( \frac{( \tilde{v}'_t )^2}{ \tilde{v}_t } + \tilde{v}_t 
\Bigr) \Bigr) dt
\\
&\le
\frac{T}{h} +
\Gamma_T \int_0^T
\Bigl( 4 V_0
\frac{ u_t^2 (u'_t)^2 }{ u_t^2} + V_0 u_t^2  \Bigr) dt
\\
&\le
\frac{T}{h} + 4 \Gamma_T V_0 \int_0^T ( (u'_t)^2  +  u_t^2 ) dt
\\
&\le
\Bigl( \frac{\yBar}{V_0} \Bigr)^{\frac12} \frac{T}{\tanh(T/2)} +
4 \Gamma_T V_0 \int_0^T ( (u'_t)^2  +  u_t^2 ) dt
\end{aligned}
\]
and we just have to integrate the expressions for $u_t$ and $u'_t$ over $[0,T]$.
Since we are interested in an upper bound for the integral, we simplify the computations using
\[
u_t \le \Bigl( \frac{\yBar}{V_0} \Bigr)^{\frac12}
\Bigl( \frac{\sinh(t/2)}{\sinh(T/2)} + 1\Bigr),
\hspace{10mm}
u'_t \le \frac12 \Bigl( \frac{\yBar}{V_0} \Bigr)^{\frac12} \frac{\cosh(t/2)}{\sinh(T/2)}.
\]
Hence, setting
\[
\begin{aligned}
&c^{(1)}_T = \int_0^T \Bigl( \frac{\sinh(t/2)}{\sinh(T/2)} + 1\Bigr)^2 dt;
\hspace{4mm}
c^{(2)}_T = \frac14 \frac1{\sinh(T/2)^2}\int_0^T \cosh(t/2)^2 dt
\\
&\tilde{c}_T = 2 \Bigl( \frac{T}{\tanh(T/2)} + 4 V_0 (c^{(1)}_T + c^{(2)}_T ) \Bigr) 
\end{aligned}
\]
we obtain that, if $y$ satisfies (\ref{e:yCond}) so that in particular
$(\frac{\yBar}{V_0})^{1/2} < \frac{\yBar}{V_0} < 2 |y|$,
\[
\int_0^T F_{\tilde{x},\tilde{v},\tilde{R}} (t) dt \le \tilde{c}_T \Gamma_T |y|
= \tilde{c}_T \Gamma_T |y|.
\]
We remark that have $c^{(1)}_T \le \int_0^T 4 dt \le 4 T$,
$c^{(2)}_T \le \frac14 \frac{T}{\tanh(\frac T2)^2} \le
\frac 1 T + T$ and $ \frac{T}{\tanh(T/2)} \le T + 1$, hence
$\tilde{c}_T \le 4 (20 V_0+1) ( \frac 1 T + T)$ for a positive constant $c$
depending on $V_0$.
On the other hand, recalling the expression of $\Gamma_T$ from (\ref{e:Gamma}),
we have $\Gamma_T \le \frac{\Gamma}{\rhop^2} e^{C_2 T^2}$ for a 
positive constant $\Gamma \ge 1$ depending on $V_0, k, \etaUnder, \sigmaUnder, \etaBar,
\sigmaBar$ but not on $\rhop$ or $T$, hence
$\tilde{c}_T \Gamma_T \le 4 (20 V_0+1) \frac{\Gamma}{\rhop^2} ( \frac 1 T + T) e^{C_2 T^2}
\le \frac{c^*}{\rhop^2} ( \frac 1 T + 1) e^{c^* T^2}$ with $c^* \ge 1$.
Now, by (\ref{e:ourRho}), the constant $Q(\mu)$ in (\ref{e:Q}) is given by
\[
Q(\mu) =
\frac{\gamma^2 q}{\rhop^4 \lambda^2} \ln \frac{\gamma q}{\rhop^2 \lambda}
\]
with $q = 8^{12} e^2 4^{73}$.
Eventually multiplying the constant $c^*$ by $ \frac{\gamma^2 q}{\lambda^2}
\ln \frac{\gamma q}{\lambda}$, we conclude that
\begin{multline}
\exp \Bigl( - Q(\mu) \Bigl( 1 + \int_0^T F_{\tilde{x},\tilde{v},\tilde{R}} (t) dt \Bigr) \Bigr) \le
\\
\exp \Bigl( - 2 c^* \Bigl( \frac 1 T + 1\Bigr) e^{c^* T^2} \frac1{\rhop^6 }
\Bigl( \ln \Bigl( \frac{1}{\rhop} \Bigr) + 1 \Bigr) \times |y| \Bigr)
\end{multline}
for every $y$ satisfying (\ref{e:yCond}).
Using (\ref{e:probaTubeEstim}) and the definition of $\psi$ in (\ref{e:psi}), the proof is completed.
}
\endproof

\bigskip

We now prove Corollary \ref{c:cdf}.

\proof[Proof of Corollary \ref{c:cdf}]{
We consider $y \in R^*$ with $|y| > (1 - V_0)/ 2$ and we now define $\yBar = 2 |y| + V_0$ and consider $\tilde{x}_t, \tilde{v}_t, \tilde{R}_t$ as in (\ref{e:curves}).
We remark that $\tilde{R}_T = \frac{1}{2} \sqrt{V_0\wedge 1} \sqrt{ 2 |y| + V_0} \le \frac{y}{2}$ if $|y|$ is larger that the larger root of $|y|^2 V_0 - 2 |y| - V_0$.
This holds in particular if $|y| > 2 (V_0 \vee 1) (1 + V_0)$ as in (\ref{e:yCond2}).
If $y > 0$, we write
\[
\begin{aligned}
&\Prob (X_T > y) \ge \Prob \Bigl(| X_T - 2 y| \le \frac{y}{2} \Bigr)
\ge \Prob (| X_T - 2 y| \le \tilde{R}_T)
\ge \Prob (| (X_T,V_T) - (2 y,\yBar)| \le \tilde{R}_T)
\end{aligned}
\]
and
\[
\begin{aligned}
\Prob (X_T < - y) \ge \Prob \Bigl( | X_T - ( -2 y) | \le \frac{|y|}{2} \Bigr)
&\ge \Prob (| X_T - (-2 y)| \le \tilde{R}_T) \\
&\ge \Prob (| (X_T,V_T) - (-2 y,\yBar)| \le \tilde{R}_T)
\end{aligned}
\]
and in both cases the last term is larger than $\Prob (| (X_t,V_t) - (\tilde{x}_t,\tilde{v}_t)| \le \tilde{R}_t,
t \in [0,T])$.
Theorem \ref{t:lowerTube} then yields estimate (\ref{e:cdf}).
To prove (\ref{e:momExpl}), we shall first show that if $\esp[ e^{p X_T}] < \infty$,
then $p \le c_T \psi(\rhop)$.
Indeed, it is sufficient to observe that if $\esp[ e^{p X_T}] = C < \infty$, $p > 0$, then $\Prob(X_T > y) \le C e^{-p \times y}$ for all $y > 0$ by Markov' inequality:
\be \label{e:upperTails}
\Prob(X_T > y) = \Prob(e^{p X_T} > e^{p y}) \le e^{-p \times  y} \esp[ e^{p X_T}].
\ee
Since (\ref{e:cdf}) and (\ref{e:upperTails}) hold simultaneously
for all $y$ from a certain range on, clearly this implies $p \le
c_T \psi(\rhop)$.
With the same argument and using the estimate for $\Prob(X_T < -y)$, one shows
that if $\esp[ e^{-q X_T}] < \infty$, $q >0$, then $q \le c_T \psi(\rhop)$.	
Finally, the estimate (\ref{e:implVolAs}) on the implied volatility is a direct consequence of moment formula (\ref{e:momForm}) and of (\ref{e:momExpl}), recalling that the function $\varphi$ is decreasing.
}
\endproof

\section{Proof of results in \ref{s:density}} \label{s:proofs2}

\noindent
We introduce some compact notation that will be used throughout this section.
For $t, s$ with $0 \le t < s \le T$ and $x_1, v_1 \in \R \times [0, \infty)$ 
we denote $(X_u^{t,x_1}, V^{t,v_1}_u; t \le u \le s)$ the solution of
(\ref{e:LSV1})-(\ref{e:LSV2}) on $[t,s]$ with initial conditions
$X_t = x_1$ and $V_t = v_1$. 
We denote $Y_u^{t,x_1,v_1}$ the couple $(X_u^{x_1}, V^{v_1}_u)$ and
\be \label{e:strLines}
\begin{aligned}
&x_u^{x_1, x_2} = x_1 + \frac{x_2-x_1}{s-t}(u-t), \ \ \ u \in [t,s]
\\
&v_u^{v_1, v_2} = v_1 + \frac{v_2-v_1}{s-t}(u-t), \ \ \ u \in [t,s]
\end{aligned}
\ee
the line segments between $(t,x_1)$,$(s,x_2)$ and $(t,v_1)$,$(s,v_2)$
respectively.
For $y \neq 0$ and a couple of radii $R_1, R_2$ with $0 < R_2 \le R_1 \le |y|$, we define
\[
\begin{aligned}
&A_{t, s}^{x_1, v_1}(y,R_2) :=
\bigl\{ |Y_u^{t,x_1,v_1} - (x^{x_1,y}_u, v_u^{v_1, y+|V_0|})| \le R_2,
u \in [t, s] \bigr\};
\\
\\
&p_{t,s}(y,R_1,R_2) = \inf_{(x_1, v_1) \in B_{R_1} (y,|y|+V_0)}
\Prob\bigl( A_{t, s}^{x_1, v_1} (y,R_2) \bigr).
\end{aligned}
\]
Moreover, we set
\be \label{e:epsilon0}
\epsilon_0 = \frac{ \rhop \etaUnder \: \sigmaUnder }{4 \sqrt{2} \rho \: \etaBar}
\wedge 1
\ee
with $\epsilon_0 = 1$ if $\rho = 0$, and
\be \label{e:delta0}
\delta_0 = \frac{\epsilon_0^2 q}{160 K^2} \wedge \frac T2
; \hspace{10mm}
q = \Prob\Bigl(\sup_{u \le 1} |b_u| \le \frac{\epsilon_0}{4 \sqrt{2} \mspace{3mu} \sigmaBar}\Bigr)
\ee
where $(b_u, u \ge 0)$ is a standard Brownian motion under $\Prob$.
The following lemma provides some estimates that will be used in the proof of 
Proposition \ref{p:smallBalls} and Theorem \ref{t:density}.

\begin{lem} \label{l:infInterval}
Let $y \in \R$ with $|y|> 16$ and $R_1, R_2$ with $0 < R_2 \le R_1 \le \sqrt{|y|}$.
Assume (R').
Then, for any $0 \le t < s \le T$, 
\be \label{e:incr1}
p_{t,s}(y,R_1,R_2)
\ge \exp \Bigl( - c_T \psi(\rhop)
\Bigl( \frac{R_1^2}{(s-t)y} + \frac{y^2}{R_2^2}(s-t) \Bigr) \Bigr)
\ee
where $c_T$ is the constant defined in (\ref{e:timeConst}).
Moreover, if $y > 0$, for any $t >0$ and any $0 < \delta < \frac{ \delta_0 }{y}
\wedge t$ we have
\begin{multline} \label{e:incr2}
\inf_{v \in B_{\epsilon_0 \sqrt{\delta y}/2}(y)}
\Prob \Bigl( | V_s^{t, v} - y | < \epsilon_0 \sqrt{\delta y}, t-\delta \le s \le t;
\\
\Bigl| \int_{t-\delta}^t ( \sigma(u, V^{t, v}_u) - \sigma(t-\delta,V^{t, v}_{t-\delta}) ) \sqrt{V_u^{t,v}} dW^1_u \Bigr|
\le \epsilon_0 \sqrt{\delta y} \Bigr) \ge \frac12 q.
\end{multline}
\end{lem}

\bigskip

The proof of this lemma is not particularly enlightening for the rest of our study, hence we postpone it to Appendix \ref{a:app2}.
Here we give the proof of Proposition \ref{p:smallBalls}.

\bigskip

\proof[Proof (of Proposition \ref{p:smallBalls})]
{\emph{Step 1.}
We consider $R_1, R_2$ with $0 < R_1 < R_2 \le \sqrt{|y|}$
and $\frac T2 \le t < s \le T$.
We have $\{ Y_s \in B_{R_2} (y,|y|+V_0) \} \supset
\{ Y_t \in B_{R_1} (y,|y|+V_0) \} \cap \{ Y_u \in B_{R_2} (x^{X_t,y}_u,v^{V_t,|y|+V_0}_u), t < u \le s \}$.
Hence, applying Markov property for the process $Y$
\be \label{e:Markov}
\begin{aligned}
\Prob( Y_s \in B_{R_2} (y,|y|+V_0) )
&\ge \Prob( \{ Y_t \in B_{R_1} (y,|y|+V_0) \} \cap \{ Y_u \in B_{R_2} (x^{X_t,u}_u,v^{V_t,|y|+V_0}_u), t < u \le s \} )
\\
&= \esp \Bigl[ 1_{ \{ Y_t \in B_{R_1} (y,|y|+V_0) \} }
\esp \Bigl[ 1_{ \{ Y_u \in B_{R_2} (x^{X_t,y}_u,v^{V_t,|y|+V_0}_u), t < u \le s \} } | \F_t \Bigr] \Bigr]
\\
&= \esp \Bigl[
1_{  \{ Y_t \in B_{R_1} (y,|y|+V_0) \} }
\esp\Bigl[
1_{ A^{x_1,v_1}_{t,s}(y,R_2) } | Y_t = (x_1,v_1) \Bigr] \Bigr]
\\
&\ge \Prob( Y_t \in B_{R_1} (y,|y|+V_0) ) \times p_{t,s}(y,R_1,R_2).
\end{aligned}
\ee

\emph{Step 2.}
We define the time step
\[
\delta_j = \delta_j(y) = \frac{ T }{ 2 |y|^j},
\ \ \ j \ge 1.
\]
Applying Lemma (\ref{l:infInterval}), for any $j \ge 1$ we have
\be \label{e:transition}
\begin{aligned}
\inf_{\frac T2 \le t \le T-\delta_j} p_{t,t+\delta_j}(y,R^{(j-1)}(y), R^{(j)}(y)) &\ge
\exp \Bigl( - c_T \psi(\rhop)
\Bigl( \frac{ (R^{(j-1)}(y))^2 }{ \delta_j(y) |y| }
+ \frac{y^2}{ (R^{(j)}(y))^2 } \delta_j(y) \Bigr) \Bigr)
\\
&= 
\exp \Bigl( - c_T \psi(\rhop)
\Bigl( 2 \frac{ y^{2-j} }{ T |y|^{1-j} }
+ \frac{y^2}{ |y|^{1-j} } \frac{ T }{  2 |y|^j} \Bigr) \Bigr)
\\
&= 
\exp \Bigl( - 2 c_T \psi(\rhop)
\Bigl( \frac1T + T \Bigr) |y| \Bigr).
\end{aligned}
\ee
On the other hand, $\tilde{R} = \frac12 \sqrt{ (V_0 \wedge 1) (|y| + V_0) }
\le \frac12 \sqrt{|y| + V_0} \le \sqrt{ \frac{|y|} 2} \le R^{(0)}(y)$.
Applying Theorem (\ref{t:lowerTube}) on the interval $[0,t]$, we have
\be \label{e:slice}
\begin{aligned}
\inf_{\frac T2 \le t \le T} \Prob( Y_t \in B_{R^{(0)}(y)} (y,|y|+V_0) ) &\ge 
\inf_{\frac T2 \le t \le T} \Prob( Y_t \in B_{\tilde{R}} (y,|y|+V_0) )
\\
&\ge \inf_{\frac T2 \le t \le T} \exp \Bigl( -c_t \Bigl(\frac1t + 1\Bigr) \psi(\rhop) |y| \Bigr)
\\
&\ge \exp\Bigl( -2 c_T \Bigl(\frac1T + 1\Bigr) \psi(\rhop) |y| \Bigr).
\end{aligned}
\ee

\emph{Step 3.}
We fix $j \in \mathbb{N}^*$ and define
\[
t^j_k = T - \sum_{h=1}^k \delta_{j-h+1},
\ \ \ 0 \le k \le j,
\]
so that $t_0 = T$ and $t^j_{k-1}-t^j_k = \delta_{j-k+1}$ for $1 \le k \le j$.
Moreover, since $\sum_{k=1}^{\infty} \delta_k = \frac T2
\sum_{k=1}^{\infty} \frac1{ y^j} \le \frac T2 \frac1{y-1} \le \frac T2$, we have
$t^j_k \ge \frac T2$ for all $j \in \mathbb{N}^*, 1 \le k \le j$.
Repeatedly applying (\ref{e:Markov}) and (\ref{e:transition}), we get 
\[
\begin{aligned}
\Prob( Y_T \in B_{R^{(j)}}(y,|y|+V_0) ) &= \Prob( Y_{t_0} \in B_{R^{(j)}}(y,|y|+V_0) )
\\
&\ge
\Prob( Y_{t_j} \in B_{R^{(0)}}(y,|y|+V_0) )
\prod_{k=1}^j p_{t^j_k, t^j_{k-1}} (y, R^{(j-k)}(y), R^{(j-k+1)}(y))
\\
\\
&\ge
\Prob( Y_{t_j} \in B_{R^{(0)}}(y,|y|+V_0) ) \times
\exp\Bigl( -2 j  c_T \Bigl( \frac1T + T \Bigr) \psi(\rhop) |y| \Bigr)
\\
&\ge \exp\Bigl( -2 (j+1) c_T \Bigl(\frac1T + T \vee 1\Bigr) \psi(\rhop) |y| \Bigr)
\end{aligned}
\]
and in the last step we have applied (\ref{e:slice}).
Using the expression for the the constant $c_T$ given in (\ref{e:timeConst}),
we have $c_T \Bigl(\frac1T + T \vee 1\Bigr) \le c^* \Bigl(\frac1T + 1\Bigr) e^{c^* T^2} \Bigl(\frac1T + T \vee 1\Bigr) \le 2 c^* \Bigl(\frac1{T^2} + 1\Bigr) e^{(c^* +1)T^2}$ and (\ref{e:smallBalls}) is proved.
}
\endproof

\bigskip

Let us go back Theorem \ref{t:density}.
To lower bound the density of $X_T$ we follow the approach of \cite{BalCar}, section 5.
The idea is to treat $X_T$ as a random variable of the form
\[
F = x + G + R,
\]
where $x \in \R$, $R \in \mathbb{D}^{2,\infty}$ and $G$ is a Wiener integral
$G = \sum_{j=1,2} \int_0^T h_j(t) dW^j_t$, with $h_j: [0,\infty) \to \R$ deterministic.
Here $\mathbb{D}^{2,\infty}$ denotes the space of the random variables which
are two times Malliavin differentiable in $L^{p}$ for every $p \ge 2$
(we refer to \cite{Nual06} for the notation and for a general presentation of Malliavin calculus; see also Appendix \ref{a:app3} for a reminder of the main elements of this theory).
Remark that $G$ is a centered Gaussian random variable with variance $\Delta =
\sum_{j} \int_0^T h_j(t)^2 dt > 0$.
Let $g_{\Delta}(y) = \frac1{\sqrt{2 \pi \Delta}} \exp( -\frac{y^2}{2 \Delta})$ denote the density of $G$ and $||R||_{2,p}$ the stochastic Sobolev norm of $R$
of order two.
Our starting point is the following result due to Bally and Caramellino in \cite{BalCar}, which we restate here in a form suitable for our purposes.

\begin{prop}[Proposition 8 in \cite{BalCar}] \label{p:genDens}
If the law of $F$ has a density $p_F$, then for any $f \in C_b(\R)$ one has
\be \label{e:genDens}
\int_{R} f(y) p_F(y) \ge \int_{R} f(y) \bigl( g_{\Delta}(y-x) - \epsilon(\Delta, R) \bigr) dy,
\ee
with
\[
\epsilon(\Delta, R) = \frac{C^*}{\sqrt{\Delta}}\bigl(1+||R_{\Delta}||_{2,q^*}\bigr)^{l^*}
||R_{\Delta}||_{2,q^*}
\]
where $R_{\Delta} = R / \sqrt{\Delta}$ and $C^*,q^*,l^*$ are universal constants.
\end{prop}

\proof{
Using point i) of Proposition 8 in \cite{BalCar}, we know that there exists a probability measure  $\overline{\Prob}$ on $(\Omega, \F)$ such that 
$\frac{ d \overline{\Prob}}{d\Prob} \le 1$ and the law of $F$ under $\overline{\Prob}$ is absolutely continuous with respect to the Lebesgue measure.
Again according to \cite{BalCar}, the associated density $\overline{p}_F$ satisfies
\[
\sup_{y \in \R} |\overline{p}_F(y) - g_{\Delta}(y-x)| \le \epsilon(\Delta, R)
\]
for the given $\epsilon(\Delta, R)$.
(We refer to \cite{BalCar} for the explicit construction of the probability $\overline{\Prob}$).
Then, for any $f \in C_b(\R)$ we have
\[
\int_{R} f(y) p_F(y) = \esp[ f(F) ] \ge
\esp \Bigl[ f(F) \frac{ d \overline{\Prob}}{d\Prob} \Big]
\ge \int_{R} f(y) \bigl( g_{\Delta}(y-x) - \epsilon(\Delta, R) \bigr) dy
\]
which proves (\ref{e:genDens}).
}

\begin{rem}
\emph{If the density $p_F$ is continuous, then (\ref{e:genDens}) implies $p_F(y) \ge g_{\Delta}(y-x) - \epsilon(\Delta, R)$ for all $y \in \R$.}
\end{rem}

\begin{rem} \label{r:condMallRem}
\emph{We shall use conditional calculus in order to prove Theorem \ref{t:density} :
in particular, we will work with Malliavin derivatives
only with respect to the Brownian noise $W_t, t \in [T-\delta,T]$, and consider conditional
expectations with respect to $\mathcal{F}_{T-\delta}$, for a $\delta < T$.
This allows us to gain a free parameter $\delta$ in (\ref{e:genDens}) that we can eventually optimize,
and this feature turns out to be crucial in our analysis (cf. Propositions \ref{p:gauss} and \ref{p:reminderEstim} hereafter).
The use of conditional Malliavin calculus in order to derive lower bounds for the density of a random variable is not new and has been employed by, among others, \cite{KH}, \cite{BalLowBounds} and
\cite{Eul2art}.
In our framework, we face some supplementary difficulties.
Let us point them out: first, to estimate the marginal density of $X_T$ we have to separately
estimate the whole path of the stochastic volatility $V$ up to time $T$.
This was the motivation of estimate (\ref{e:incr2}).
Second, in order to manipulate the Sobolev norms of $R$ we need 
all the involved random variables to be smooth in Malliavin sense,
but this is not guaranteed in our framework due to the presence of the non-Lipschitz
square-root coefficients in (\ref{e:LSV1}) and (\ref{e:LSV2}).
This is why we introduce a regularization of the coefficients of the SDE,
as we do hereafter.
}
\end{rem}

\bigskip

Let us implement what stated in Remark \ref{r:condMallRem}.	 
We consider the case of positive $y$
in Theorem \ref{t:density}: the case of negative $y$ is proven in the analogous manner.
We assume $y > 2$ and introduce two parameters $\delta > 0$ and $l \in \mathbb{N}$
such that:
\be \label{e:deltaL}
\delta < \frac{\delta_0}{y^2};
\hspace{10mm}
\frac1{y^l} < \frac12 \epsilon_0 \rhop \etaUnder \sqrt{y \delta}
\ee
and $\epsilon_0, \delta_0$ as defined in (\ref{e:epsilon0}) and (\ref{e:delta0}).
We remark that for such a value of $\delta$ we have
$\epsilon_0 \sqrt{\delta y} < \epsilon_0 \sqrt{\delta_0} < 1$.
Then, we consider a truncation function $\psi \in C^{\infty}_b(\R, \R)$ such that $\psi(x) = x$ for
$|x - y| \le 1 $, $\psi(x) = y-\frac32$ for $x \le y-2$ and $\psi(x) = y+\frac32$ for $x \ge y+2$.
$\psi$ can be defined in such a way that $|\psi|_0 \le y+\frac32 \le 2 y$ and
$\sum_{j=1}^ k \sup_{x \in \R} |\psi^{(j)} (x)| \le 2^{\frac{k(k-1)}2}$.
We define the sets
\[
A_{\delta, l}(X,V) = \{ |X_{T-\delta} - y | < \frac1{y^l}, | V_{T-\delta} - (y+V_0) | < \frac{\epsilon_0}2
\sqrt{y \delta} \}
\]
and
\begin{multline*} 
\overline{A}_{\delta}(V) = \Bigl\{ | V_s - y | < \epsilon_0 \sqrt{\delta y},
T-\delta < s \le T;
\\
\Bigl| \int_{T-\delta}^T ( \sigma(u, V_u)-\sigma(T-\delta,V_{T-\delta}) )
\sqrt{V_u} dW^1_u \Bigr| \le \epsilon_0 \sqrt{\delta y} \Bigr\},
\end{multline*}
and denote
\[
A_{\delta,l} = A_{\delta,l}(X,V) := A_{\delta,l}(X,V) \cap \overline{A}_{\delta}(V).
\]
Finally, we consider $(\Xbar_t, \Vbar_t; T-\delta \le t \le T)$ the (unique strong) solution to the equation
\be
\begin{aligned} \label{e:truncate}
\Xbar_t &= X_{T-\delta} -\frac12 \int_{T-\delta}^t \eta(s, \Xbar_s)^2 \psi(\Vbar_s) ds
        + \int_{T-\delta}^t \eta(s,\Xbar_s) \sqrt{\psi(\Vbar_s)} (\rho dW^1_s + \rhop dW^2_s), 
\\
\\
\Vbar_t &= V_{T-\delta} + \int_{\Td}^t \beta(s, \Vbar_s ) ds + \int_{\Td}^t \sigma(s,\Vbar_s)
\sqrt{ \psi(\Vbar_s)  } dW^1_s.
\end{aligned}
\ee
We remark that on the set $A_{\delta,l}$, $\psi(V_t) = V_t$ for all $t \in [T-\delta, T]$. 
Hence, since pathwise uniqueness holds for (\ref{e:truncate}), we have
$(X_t, V_t) (\omega) = (\Xbar_t, \Vbar_t) (\omega)$ for $(t,\omega) \in [T-\delta, T] \times
A_{\delta,l}$ and in particular $A_{\delta,l} = A_{\delta,l}(X,V) \subset A_{\delta,l}(\Xbar, \Vbar)$.
Under hypothesis (R'), $\Xbar_t$ and $\Vbar_t$ belong to the space $\mathbb{D}^{2,p}$
associated to $(W^1_t, W^2_t), t \in [T-\delta, T]$, for all $p > 1$.
\\
We decompose the random variable $\Xbar_T$ in the following way:
\[
\Xbar_T = G_0 + G + R, 
\]
where
\be \label{e:decomp}
\begin{aligned} 
&G_0 = X_{T-\delta} + \rho \eta(T-\delta, X_{T-\delta})
\int_{T-\delta}^T \sqrt{\psi( \Vbar_t) } d W^1_t 
\\
&G = \rhop \eta(T-\delta, X_{T-\delta})
\int_{T-\delta}^T \sqrt{\psi(\Vbar_t)} d W^2_t 
\\ 
&R = -\frac12 \int_{T-\delta}^T \eta(t, \Xbar_t)^2 \psi(\Vbar_t) dt
\\ 
&\hphantom{R = -\frac12 \int_{T-\delta}^T \eta(t, \Xbar_t)^2}
+ \int_{T-\delta}^T
( \eta(t, \Xbar_t)-\eta(T-\delta, X_{T-\delta}) ) \sqrt{\psi(\Vbar_t)} (\rho dW^1_t + \rhop dW^2_t).
\end{aligned}
\ee
Conditional to $\F_{T-\delta}\vee \F^1_T$, the random variable $G$ is a centered Gaussian with
variance $I = \rhop^2 \eta(T-\delta, X_{T-\delta})^2 \int_{T-\delta}^T 
\psi(\Vbar_t) dt $.
By the definition of $\psi$, we have $I
\ge \rhop^2 \etaUnder^2 \int_{T-\delta}^T (y-\frac32) dt \ge \frac12 \rhop^2 \etaUnder^2 y \delta$.
Similarly, we can see that an upper bound for $I$ is given by $2 \rhop^2 \etaBar^2 y \delta$, hence
\be \label{e:lowVar}
\Delta \le
I = Var( G | \F_{T-\delta}\vee \F^1_T)
\le
a \Delta
\ee
with
\[
\Delta = \frac12 \rhop^2 \etaUnder^2 y \delta, \hspace{10mm}
a = 4  \frac{\etaBar^2}{ \etaUnder^2}.
\]
Using (\ref{e:lowVar}) and Lemma (\ref{l:infInterval}), we can prove the following statement (which is the analogous of Lemma 5
in \cite{BalLowBounds}):

\begin{prop} \label{p:gauss}
Let $g(\cdot|F_{T-\delta} \vee \F^1_T)$ denote the density of $\mspace{4mu} G$
conditional to $F_{T-\delta} \vee \F^1_T$.
Then, for any $y,\delta,l$ satisfying (\ref{e:deltaL}),
\be \label{e:gauss}
g( y - G_0 | F_{T-\delta} \vee \F^1_T) \ge
\frac1{ \rhop \etaBar e \sqrt{4 \pi \delta y}} \hspace{8mm} \text{on the set $A_{\delta,l}$}.
\ee 
\end{prop}

\proof{
Recall that $I = \rhop^2 \eta(T-\delta, X_{T-\delta})^2 \int_{T-\delta}^T \psi(\Vbar_t) dt$.
Moreover, let us set $J = \rho \eta(T-\delta, X_{T-\delta}) \int_{T-\delta}^T \sqrt{\psi(\Vbar_t)} d W^1_t$ (so that $G_0 = X_{T-\delta} +J$).
Then
\be \label{e:condExpect}
g(y - G_0 | F_{T-\delta} \vee \F^1_T)
= \frac1{\sqrt{2 \pi I}} \exp \Bigl( -\frac1{2 I} (y - (X_{T-\delta}+J))^2 \Bigr).
\ee
Since $I \ge \Delta$ and $| y - X_{T-\delta} |\le \frac1{y^l}$ on $A_{\delta,l}$, on this set we have
\[
\frac{ | y - X_{T-\delta} | }{\sqrt{I}} \le \frac{ \frac1{y^l} }{\sqrt{\Delta}} \le 1 
\]
where the last inequality holds because of (\ref{e:deltaL}).
Now, using equation (\ref{e:truncate}) for $\Vbar$,
\begin{multline*}
\sigma(T-\delta, \Vbar_{T-\delta} ) \int_{T-\delta}^T \sqrt{\psi(\Vbar_t)} d W^1_t
= \Vbar_T - \Vbar_{T-\delta} - \int_{T-\delta}^T \beta(t,\Vbar_t) dt
\\
-\int_{T-\delta}^T ( \sigma(t,\Vbar_t)-\sigma(T-\delta,\Vbar_{T-\delta})) \sqrt{\psi(\Vbar_t)} d W^1_t,
\end{multline*}
hence, on the set $A_{\delta,l}$ 
\[
\begin{aligned}
\Bigl| \int_{T-\delta}^T \sqrt{\psi(\Vbar_t)} d W^1_t \Bigr|
&\le \frac1{\sigmaUnder} \Bigl( |\Vbar_T - \Vbar_{T-\delta}|
+ \int_{T-\delta}^T | \beta(t,\Vbar_t) | dt
\\
&\quad \quad + \Bigl| \int_{T-\delta}^T ( \sigma(t,\Vbar_t)-\sigma(T-\delta,\Vbar_{T-\delta})) \sqrt{\psi(\Vbar_t)} d W^1_t \Bigr| \Bigr)
\\
&\le \frac1{\sigmaUnder} \Bigl(
2 \epsilon_0 \sqrt{y \delta} + K (1 + y + \epsilon_0 \sqrt{y \delta}) \delta
+ \epsilon_0 \sqrt{y \delta} \Bigr)
\\
&\le \frac1{\sigmaUnder} (2 \epsilon_0 +  2 K \sqrt{y \delta} + \epsilon_0) \sqrt{y \delta}
\le \frac{4 \epsilon_0 }{\sigmaUnder}\sqrt{y \delta}
\end{aligned}
\]
and the two last inequality are obtained using
$ K (1 + y + \epsilon_0 \sqrt{y \delta}) \delta < K (1 + y + 1) \delta < 2 K y \delta$, then $2 K \sqrt{y \delta} \le 2 K \sqrt{\delta_0} \le \epsilon_0$ after (\ref{e:delta0}).
The previous estimate yields 
$|J| \le \frac{4 \epsilon_0 \rho \etaBar}{\sigmaUnder}\sqrt{y \delta}$, hence
\[
\frac{|J|}{\sqrt I} \le \frac{|J|}{\sqrt{\Delta}} \le \epsilon_0 \frac{ 4 \sqrt2 \rho \etaBar }{ \rhop \etaUnder \sigmaUnder} 
\le 1
\]
and the last inequality holds after (\ref{e:epsilon0}). 
Finally, for the exponential term in (\ref{e:condExpect}) we have
\be \label{e:exp}
\exp \Bigl( -\frac1{2 I}
\Bigl( y - (X_{T-\delta} - J) \Bigr)^2 \Bigr) \ge e^{-\frac12 (1+1)^2} \ge
e^{-2}
\ee
on the set $A_{\delta,l}$.
Since $I \le a \Delta = 2 \rhop^2 \etaBar y \delta$, (\ref{e:exp}) yields
(\ref{e:gauss}).
}
\endproof

\noindent
The second result we need in order to prove Theorem \ref{t:density} is an estimation of the reminder $R$.
Let $R_{\Delta} := R / \sqrt{\Delta}$ as in Proposition \ref{p:genDens}.

\begin{prop} \label{p:reminderEstim}
Let $y,\delta,l$ satisfy (\ref{e:deltaL}). Then, for every $p > 1$ there exists a positive constant $c_p$ such that
\be \label{e:reminderEstim}
|| \RD ||_{T-\delta,\delta,2,p} \le \frac{c_p}{\rhop \etaUnder} \sqrt{ \delta y^{31} }
\times e^{c_p T^p}
\hspace{8mm} \text{on the set $A_{\delta,l}$}.
\ee
The constant $c_p$ depends also on $K$ but not on the other model parameters.
\end{prop}

\begin{rem}
\emph{
Similar estimates (with different powers of $\delta$ and $y$) could be obtained for
$|| \RD ||_{T-\delta,\delta,k,p}$, $k > 2$, under the corresponding regularity assumptions
on the coefficients $\eta, \beta$ and $\sigma$.
}
\end{rem}

We can now prove Theorem \ref{t:density}.
\vspace{1mm}

\proof[Proof (of Theorem \ref{t:density})]
{
We make an explicit choice of $\delta$ and $l$:
\be \label{e:deltaChoice}
\delta = \delta_0 \Theta_T^2 \times y^{-31};  \hspace{8mm} l = 16; \hspace{8mm}
\text{with} \hspace{4mm} \Theta_T = \frac{\rhop \etaUnder^2 2^{-l^*-5/2}}{\sqrt{\pi} e \etaBar C^* c_p}
e^{-2 c_p T^p}
\ee
where $C^*$ and $c_p$ are the constant appearing in, respectively, (\ref{e:genDens}) and (\ref{e:reminderEstim}).
Conditions (\ref{e:deltaL}) are satisfied as soon as
\[
y > M_T \hspace{4mm} \text{with} \hspace{4mm} M_T = 2/(\epsilon_0 \rhop \etaUnder
\sqrt{\delta_0} \Theta_T).
\]
We now apply Proposition \ref{p:genDens} to the law of $\Xbar_T$ conditional to $\overline{\F} = \F_{T-\delta} \vee \F^1_T$.
Let $p_{\Xbar_T} (\cdot | \overline{\F})$ (resp. $g(\cdot|\overline{\F})$) denote the conditional density of $\Xbar_T$ (resp. $G$), by Prop. \ref{p:genDens} we
have
\[
\begin{aligned}
\int_{y>M_T} f(y) p_{\Xbar_T} (y | \overline{\F}) dy &\ge 
\int_{y>M_T} f(y) \bigl( g(y - G_0|\overline{\F}) - \epsilon(\Delta,R) \bigr) dy
\\
&\ge 
\int_{y>M_T} f(y) \Bigl( \frac1{\rhop \etaBar e \sqrt{4 \pi \delta y}} - \epsilon(\Delta,R) \Bigr) dy
\end{aligned}
\]
with
\[
\begin{aligned}
\epsilon(\Delta,R) &= \frac{C^*}{\sqrt{\Delta}}
\bigl( 1 + \esp[ ||R_{\Delta}||_{2,q^*} | \overline{\F}] \bigr)^{l^*} \esp[ ||R_{\Delta}||_{2,q^*} | \overline{\F} ]
\\
&= \frac{C^*}{\sqrt{\Delta}}
\bigl( 1 + \esp[ ||R_{\Delta}||_{T-\delta,\delta,2,q^*} | \F^1_T ] \bigr)^{l^*} \esp[ ||R_{\Delta}||_{T-\delta,\delta,} | \F^1_T ].
\end{aligned}
\]
For the given value of $\delta$, $||R_{\Delta}||_{T-\delta,\delta,2,q^*} < 1$ on the set $A_{\delta,l}$, hence by Prop. \ref{p:reminderEstim} $\epsilon(\Delta,R)$ is bounded by
\[
\epsilon(\Delta,R) \le \frac{C^* c_p}{ \rhop^2 \etaUnder^2  } \times 2^{l*+1/2} \times
\sqrt{\delta y^{31}} e^{C_p T^p}
\]
on the set $A_{\delta,l}$.
The value of $\delta$ in (\ref{e:deltaChoice}) is chosen in such a way that
the right hand side in this last estimate is smaller than
$\frac12 \times \frac1{\rhop \etaBar e \sqrt{4 \pi \delta y}}$, hence
\[
\begin{aligned}
\int_{y>M_T} f(y) p_{\Xbar_T} (y | \overline{\F}) dy \ge
\int_{y>M_T} \frac{f(y)}{\rhop \etaBar e \sqrt{4 \pi \delta y}} dy =
\int_{y>M_T} \frac{f(y)}{\rhop \etaBar e \sqrt{4 \pi \delta_0} \Theta_T } y^{15} dy
\end{aligned}
\]
on the set $A_{\delta,l}$.

Let us now estimate the probability of the set $A_{\delta,l}$.
Since $\frac1{y^l} < \frac{\epsilon_0}2 \sqrt{y\delta}$ by condition (\ref{e:deltaL}),
then $A_{T-\delta} \supset \{ (X_{T-\delta}, V_{T-\delta} ) \in B_{\frac1{y^l}}(y,y+V_0) \}$.
Hence, we apply Prop. \ref{p:smallBalls} for $j = 2l+1$ and obtain
\[
\begin{aligned}
\Prob( A_{T-\delta} ) \ge \Prob\bigl( ( X_{T-\delta}, V_{T-\delta} ) \in B_{\frac1{y^l}}(y,y+V_0) \bigr)
&\ge \exp \Bigl( -(2l+2) d_{T-\delta} \psi(\rhop) y \Bigr)
\end{aligned}
\]
Since $2l+2= 34$ and $\frac1{(T-\delta)^2} < \frac4{T^2}$, then
$(2l+2) d_{T-\delta} \le 136 c^* \Bigl( \frac1{T^2} + 1 \Bigr) e^{(c^*+1)T} := e_T$
and $\Prob( A_{T-\delta} ) \ge \exp ( - e_{T} \psi(\rhop) y )$.
Applying Markov property for the process $V$ and (\ref{e:incr2}) in Lemma \ref{l:infInterval}, it is easy to see that
\be \label{e:probA}
\Prob(A_{\delta,l})
\ge \frac q2 \Prob(A_{T-\delta}) \ge \frac q2 \exp \Bigl( -e_{T} \psi(\rhop) y \Bigr) .
\ee

Finally, let us denote $p_{X_T}(\cdot|\overline{\F})$ the density of $X_T$ conditional
to $\overline{\F}$.
We have
\[
\begin{aligned}
\int_{\R} f(y) p_{X_T}(y) dy &=
\int_{\R} f(y) \esp[ p_{X_T}(y|\overline{\F})] dy
\\
&\ge\int_{\R} f(y) \esp[ p_{X_T}(y|\overline{\F}) 1_{A_{\delta,l}}] dy
\\
&= \int_{\R} f(y) \esp[ p_{\Xbar_T}(y|\overline{\F}) 1_{A_{\delta,l}}] dy
\\
&\ge \int_{\R} f(y) \frac{1 }{ \rhop \etaBar e \sqrt{4 \pi \delta_0} \Theta_T } y^{15} \times
\Prob(A_{\delta,l}) dy.
\end{aligned}
\]
Using estimate (\ref{e:probA}), we obtain (\ref{e:density}).
}
\endproof

\section*{Conclusion}
\noindent
We have shown that the left and right tails of the distribution of the log price
$X$ decay no faster than exponentials in local stochastic volatility models
driven by square root diffusions - namely, in the model class (\ref{e:ourClass1})-(\ref{e:ourClass2}) - no matter how the (possibly time-dependent) skew function $\eta$, the volatility drift $\beta$ and volatility of variance $\sigma$ are chosen, provided they satisfy some reasonable boundedness and linear-growth conditions - namely, conditions (R) and (G) in section \ref{s:main}. 
Together with the elementary observation that $e^{X}$ is an integrable supermartingale, this yields the ``sandwich'' estimate $e^{-c_1(t) y} \le \Prob(X_t > y) \le e^{-c_2(t) y}$ for large  values of $y$.
From the point of view of the financial modelisation, our estimate has an impact on moment explosion and, by Lee's moment formula, it turns into lower bounds on the asymptotic slopes of the implied volatility.

Our result is not limited to fixed-time marginal laws:
we have shown that the exponential lower bound actually holds for the probability that the whole trajectory of the couple $(X,V)$ remains in a ``tube'' of given deterministic radius around a given deterministic curve for all the times up to a given maturity.
This means that our main estimate can also be applied to the two-dimensional joint distribution of $X$ and $V$ and to study the law of suprema of the components of the solution to (\ref{e:ourClass1})-(\ref{e:ourClass2}).
Back to the financial level, this can eventually lead to bounds on the prices of barrier and exotic options.
We have also shown how one can apply density estimation techniques for locally-elliptic random variables on the Wiener space and Malliavin calculus tools to prove that a lower bound of the same range holds the for the density of $X$ as well.

An central subject for future work is the way to generalise these results to a wider class of models, with particular focus on allowing for different powers of $V$ in the equation for $X$ and in the one for $V$.
Our current impression is that this kind of program calls for a sharper version of the fundamental ``tube'' estimate for It\^o processes which is at the basis of our analysis.

\section*{Appendix} \label{s:app}

\subsection{Some preliminary estimates}

\begin{lem} \label{l:preliminary}
Assume \emph{(G)} and let $(X_t,V_t;0 \le t \le T)$ be two processes satisfying (\ref{e:LSV1})-(\ref{e:LSV2}). Then for every $0 \le t \le s \le T$ and every $p \ge 1$
there exist a positive constant $C_p$
such that
\be \label{e:preliminary}
\esp \Bigl[ \sup_{t \le r \le s }
\Bigl( |X_r - X_t|^{2 p} + |V_r - V_t|^{2 p} \Bigr) \Big| \mathcal{F}_t
\Bigr] \le C_p (s - t)^p \exp( C_p s^{2p} ).
\ee
$C_p$ also depends on the parameters $K, \etaBar, \sigmaBar$
given in \emph{(G)} and on $V_0$.
\end{lem}

\proof[Proof]{
Observing that both the functions $v \to \beta(t,v)$ and $v \to \sigma(t,v)\sqrt{v}$ have sub-linear growth under (G), (\ref{e:preliminary}) follows from the application of Burkholder's inequality and Gronwall's Lemma to the process $(V_t; 0 \le t \le T)$ satisfying (\ref{e:LSV1}), then to $(X_t;0 \le t \le T)$ satisfying (\ref{e:LSV2}).}

\subsection{Proof of Lemma \ref{l:infInterval}} \label{a:app2}

\proof
Estimate (\ref{e:incr1}):
Consider $(x_1, v_1) \in B_{R_1}(y,|y|+V_0)$.
On the set \\$A_{t, s}^{x_1, v_1} (x^{x_1,y}_{\cdot}, v^{v_1,|y|+V_0}_{\cdot},R_{\cdot})$, with $R_u = R_2, u \in [t,s]$,
we have $V_u > |y| - R_1 - R_2 > |y| - 2 \sqrt{|y|}$ and
$V_u < |y| + R_1 + R_2 < |y| + 2 \sqrt{|y|}$ for all $u \in [t,s]$. 
Therefore, $\frac12 |y| < V_u < 2 |y|$ for all $u \in [t,s]$, if $|y|>16$.
Using estimates (\ref{e:C}), (\ref{e:lambdGamma}), (\ref{e:L1}) and (\ref{e:L2}) it is easy to show that conditions
(\ref{e:H1}),(\ref{e:H2}) and (\ref{e:H3}) are satisfied
for the process $(X_u^{t,x_1}, V^{t,v_1}_u; t \le u \le s)$, the curves
$x^{x_1,y}_{\cdot}, v^{v_1,|y|+V_0}_{\cdot}$, the radius $R_{\cdot}$
and the \emph{constant} curves
\be 
\begin{array}{l}
c_u = c |y|; \hspace{10mm} L_u^2 = L_T |y|;
\\
\gamma_u = \gamma |y|; \hspace{9mm}
\lambda_u = \rhop^2 \lambda |y|,
\end{array}
\ee
defined for $u \in [t,s]$, where $c, L_T, \gamma, \lambda$ are the same as in Proposition \ref{p:tech}.
The derivatives $x'_{\cdot}, v'_{\cdot}$ and the radius $R_{\cdot}$ being constant, all the involved curves belong to $L(1, \infty)$.
The factor $\phi_{\lambda,\gamma}$ in (\ref{e:Q})
is still given by $\phi_{\lambda,\gamma} = \frac{\rhop^2 \lambda}{\gamma}$,
hence $Q_{\lambda,\gamma} = \frac{\gamma^2 q}{\rhop^4 \lambda^2} \ln \frac{\gamma q}{\rhop^2 \lambda}$
with $q = 8^{12} e^2$.
On the other hand, using the constant $\Gamma_T$ defined in (\ref{e:Gamma}),
we have
\[
\begin{aligned}
\int_t^s F_{x,v,R} (u) du &=
\int_t^s \Bigl( \frac{( x'_u )^2 + ( v'_u )^2}{\lambda_u}
+ 2 (c_u^2 + L_u^2 ) \Bigl( \frac{1}{\lambda_u} + \frac{1}{R_u^2}
\Bigr) \Bigr) du
\\
&\le
4 \Gamma_T \int_t^s \Bigl( \frac{R_1^2}{(s-t)^2 y}
+ y^2 \Bigl( \frac{1}{y} + \frac{1}{R_2^2} \Bigr) \Bigr) du
\\
&\le
8 \Gamma_T \Bigl( \frac{R_1^2}{(s-t) y}
+ \frac{y^2}{R_2^2} \Bigr)(s-t) \Bigr).
\end{aligned}
\]
Estimate (\ref{e:probaTubeEstim}) then tells that (\ref{e:incr1}) holds with 
the same constant $c_T$ defined in (\ref{e:timeConst}).

(\ref{e:incr2}): We fix $t >0$, $v \in B_{\epsilon_0 \sqrt{\delta y}/2}(y)$ and
write $V = V^{t, v}$ for simplicity.
We set $B = \{ | V_s - y | < \epsilon_0 \sqrt{\delta y}, t-\delta \le s \le t \}$
and $C = \Bigl\{ \Bigl| \int_{t-\delta}^t ( \sigma(u, V_u) - \sigma(t-\delta,V_{t-\delta}) ) \sqrt{V_u}
dW^1_u \Bigr| \le \epsilon_0 \sqrt{\delta y} \Bigr\}$.
Since $\epsilon_0 \sqrt{\delta y} < \epsilon_0 \sqrt{\delta_0} < 1$, on the set $B$ we have $V_s < 1 + y < 2 y$ and $|\beta(s, V_s)| < K ( 2 + y) < 2 K y$ for all $s \in [t-\delta, t]$.
Hence, on the set $B$
\[
\begin{aligned}
|V_s - v| &\le 
2 K y \delta + \Bigl| \int_{t-\delta}^s \sigma(u, V_u) \sqrt{V_u \wedge 2 y}
d W^1_u \Bigr|
\end{aligned}
\]
and $2 K y \delta < \frac{\epsilon_0}4 \sqrt{\delta y}$ because $\delta < \delta_0/y$.
Therefore
\[
\Prob(B) \ge \Prob \Bigl( \sup_{t-\delta \le s \le t} |V_s - v| \le \frac{\epsilon_0}2 \sqrt{\delta y} \Bigr) 
\ge \Prob \Bigl( \sup_{t-\delta \le s \le t} \Bigl| \int_{t-\delta}^s \sigma(u, V_u) \sqrt{V_u \wedge 2 y}
d W^1_u \Bigr| \le \frac{\epsilon_0}4 \sqrt{\delta y} \Bigr).
\]
We time-change the stochastic integral into $b_{A_s}$, where $(b_s; s \ge 0)$ is a standard Brownian motion and we denote $A_s = \int_{t-\delta}^s \sigma(u, V_u)^2 ( V_u \wedge 2 y) du$
the quadratic variation (Dubins \& Schwartz theorem, cf. Th. 3.4.6 in \cite{KS}).
Since $A_s$, $s \in [t-\delta,t]$, is uniformly bounded by $\sigmaBar^2 2 y \delta$, we have
$\sup_{t-\delta \le s \le t} |b_{A_s}| \le \sup_{ 0 \le s \le 2 \sigmaBar^2 y \delta} |b_s|$.
Using the scaling propertyof the Brownian motion $(b_{c s}; s \ge 0) \sim ( \sqrt{c} b_s; s \ge 0)$ we obtain
\[
\Prob(B) \ge \Prob \Bigl( \sup_{ 0 \le s \le 2 \sigmaBar^2 y \delta} |b_s| \le \frac{\epsilon_0}4 \sqrt{\delta y} \Bigr) = 
\Prob \Bigl( \sup_{ 0 \le s \le 1} |b_s| \le \frac{\epsilon_0}{4\sqrt2 \sigmaBar}
\Bigr) = q.
\]
The same arguments lead to
\[
\begin{aligned}
\Prob(B \cap C^c) &\le \Prob \Bigl( \Bigl| \int_{t-\delta}^t ( \sigma(u, V_u) - \sigma(t-\delta,V_{t-\delta}) ) \sqrt{V_u} dW^1_u \Bigr| > \epsilon_0 \sqrt{\delta y} ; V_u < 2y, t-\delta \le u \le t \Bigr)  
\\
&\le \Prob \Bigl( \Bigl| \int_{t-\delta}^t ( \sigma(u, V_u) - \sigma(t-\delta,V_{t-\delta}) ) \sqrt{V_u \wedge 2y} dW^1_u \Bigr| > \epsilon_0 \sqrt{\delta y} \Bigr)
\\
&= \Prob ( | \tilde{b}_{B_t} | > \epsilon_0 \sqrt{\delta y} )
\end{aligned}
\]
where $B_t = \int_{t-\delta}^t ( \sigma(u, V_u) - \sigma(t-\delta,V_{t-\delta}) )^2 ( V_u \wedge 2 y) d u$ and $(\tilde{b}_s; s \ge 0)$ is a standard Brownian motion.
Using (R), we have $B_t \le 2 K^2 ( \delta^2 + |V_u - V_{t-\delta}|^2) \cdot 2 y \cdot \delta \le
4 K^2 ( \delta^2 + 4 \epsilon_0^2 \delta y) \cdot y \delta
\le
20 K^2 \epsilon_0^2 (\delta y)^2$,
hence
\[
\begin{aligned}
\Prob(B \cap C^c) &\le \Prob \Bigl( \sup_{s \le 20 K^2 \epsilon_0^2 (\delta y)^2} | \tilde{b}_s | > \epsilon_0 \sqrt{\delta y} \Bigr)
\\
&= \Prob \Bigl( \sup_{s \le 1} | \tilde{b}_s | > \frac{ 1 }{ \sqrt{20 K^2 \delta y} } \Bigr)
\\
&\le
20 K^2 \delta y \: \esp \Bigl[ \sup_{s \le 1} | \tilde{b}_s |^2 \Bigr]
\le 80 K^2 \delta_0
\le \frac12 q
\end{aligned}
\]
and we have used Doob's inequality and the value of $\delta_0$ to get
the two last inequalities.
We conclude that
\[
\Prob(B \cap C) = \Prob(B) -\Prob(C^c \cap B) \ge q - \frac12 q = \frac12 q.
\]
\endproof

\subsection{Conditional Malliavin calculus} \label{a:app3}

\noindent
We briefly introduce the main elements of conditional Malliavin calculus.
We consider a probability space $(\Omega, \F, \Prob)$ with a filtration
$(\F_t, t \ge 0)$, and a $d$-dimensional Brownian motion $(W_t, t \ge 0)$ with
respect to this filtration.
We fix some $t>0$ and $\delta>0$.
Conditional Malliavin calculus amounts to consider the standard Malliavin derivative operators, but focusing on the derivatives with respect to $W_t, t \in [t,t+\delta]$ on the one hand, and to replace expectations with conditional expectations with respect to $\F_t$ on the other hand.
We recall the basic notation of Malliavin calculus (we refer to \cite{Nual06} for a more detailed presentation of this topic).
$\mathbb{D}^{k,p}$ denotes the space of the random variables which are $k$ times differentiable in Malliavin's sense in $L^p$.
Let $\Theta_k = \{1, \dots, d \}^k$ be the set of multi-indexes of length $k$ with
components in $\{1,\cdots,d\}$, and let $\R^{\Theta_k} = \{ (x_{\alpha})_{\alpha \in \Theta_k} : x_{\alpha} \in \R \}$.
For a measurable function $V : [0,\infty)^k \to \R^{\Theta_k}$, we define
\[
\begin{aligned}
&|V|_k^2 := \int_{[0,\infty)^k} \sum_{\alpha \in \Theta_k}
|V^{\alpha}(s_1, \dots, s_k)|^2 ds_1 \cdots ds_k \hspace{5mm} \text{and}
\\
&H_k := \{ V : [0,\infty)^k \to \R^{\Theta_k} : |V|_k^2 < \infty \}.
\end{aligned}
\]
For $F \in \mathbb{D}^{k,p}$ and $\alpha \in \Theta_k$, the derivative of $F$ of order $k$ and index $\alpha$ is $D^{k,\alpha} F$.
We denote $D^k_{s_1, \dots,s_k} F = ( D^{k,\alpha}_{s_1, \dots,s_k} F)_{\alpha \in \Theta_k}$: it is known (cf. \cite{Nual06}) that $\esp[ | D^{k} F |_k^p ] < \infty$
for all $p \ge 1$, hence $D^k F \in H_k$.

The conditional Malliavin calculus is based on the use of the following scalar product and norm: for every fixed $t,\delta > 0$ and $U,V: [t,t+\delta) \to R^{\Theta_k}$ we define
\[
\begin{aligned}
&\langle U,V \rangle_{t,\delta,k} := \int_{[t,t+\delta)^k} \sum_{\alpha \in \Theta_k}
V^{\alpha}(s_1, \dots, s_k) U^{\alpha}(s_1, \dots, s_k) ds_1 \cdots ds_k,
\\
&|V|_{t,\delta,k}^2 := \langle V,V\rangle_{t,\delta,k}^2 = \int_{[t,t+\delta)^k} \sum_{\alpha \in \Theta_k} | V^{\alpha}(s_1, \dots, s_k)|^2 ds_1 \cdots ds_k.
\end{aligned}
\]
For $F \in \mathbb{D}^{k,p}$, we define the following Sobolev norms:
\[
\begin{aligned}
&|| F ||_{t,\delta,k}^2 := \sum_{i=0}^k |D^i F|_{t,\delta,i}^2 =
\sum_{i=0}^k \sum_{\alpha \in \Theta_i} \int_{[t,t+\delta)^i}
|D^{i,\alpha}_{s_1, \cdots, s_i} F|^2 ds_1 \cdots ds_i,
\\
&|| F ||_{t,\delta,k,p}^p := \esp[ || F ||_{t,\delta,k}^p | \F_t ];
\hspace{5mm}
||| F |||_{t,\delta,k,p}^p := || F ||_{t,\delta,k,p}^p -\esp[ |F|^p | \F_t ].
\end{aligned}
\]
Notice that $|| F ||_{t,\delta,k,p}$ is not a constant (as happens in the
standard Malliavin calulus) but an $\F_t$-measurable random variable.
The standard norm $|| F ||_{k,p}$ corresponds to $|| F ||_{0,\infty,k,p}$.
Remark that by the definition of $||| F |||_{t,\delta,k,p}$, using the elementary inequality $\frac{b^{p/2}}2 \le (a^2 + b)^{p/2} \le 2^{p/2} b^{p/2}$ for positive $a, b$ and $p \ge 2$, we have
\be \label{e:tripleNorm}
\frac12 f \le ||| F |||_{t,\delta,k,p}^p \le 2^{p/2} f,
\ee
with $f = \esp\Bigl[ \Bigl( \sum_{i=1}^k \sum_{\alpha \in \Theta_i}
\int_{[t,t+\delta)^i} |D^{i,\alpha}_{s_1, \dots, s_i} F |^2 ds_1 \cdots ds_i \Bigr)^{p/2} \Bigr]$.

We will make use of the two following inequalities: first, let $F,G \in \cap_{p\ge1} \mathbb{D}^{k,p}$.
Then, for every $p\ge 1$,
\be \label{e:derivProd}
|| F G||_{t,\delta,k,p} \le k! 2^k || F ||_{t,\delta,k,2p} || G ||_{t,\delta,k,2p}.
\ee
In addition, for every $k\ge1$ there exists a constant $\mu(k)$ such that for every $\phi \in C^k_b$, every $p > 1$ and every $F \in \mathbb{D}^{k,p}$, one has
\be \label{e:chain}
||| \phi(F) |||_{t,\delta,k,p} \le \mu(k) | \phi |_{k} ||| F |||_{t,\delta,k,
2^kp},
\ee
where $|\phi |_{k} = \sum_{i=0}^k \sup_{x \in \R} |\phi^{(i)}(x)|$.
Inequality (\ref{e:chain}) is a consequence of the chain rule and of
(\ref{e:derivProd}).
The proof of (\ref{e:derivProd}) is based on some rather standard (but cumbersome) computations and can be found in \cite{BalLowBounds}, Lemma 2.5 in the Appendix.

\bigskip

Let us now consider diffusion processes.
We consider some $T >0$ and $0 < \delta < 1 \wedge T$ and $(Y_t; t \in [T-\delta,T])$ the unique strong solution to the equation
\be \label{e:diffusion}
Y_t =
Y_{T-\delta} + \int_{T-\delta}^t B(s,Y_s)ds + \sum_{j=1}^d \int_{T-\delta}^t A_j(s,Y_s)dW^j_s, \ \ \ T-\delta \le t \le T,
\ee
where $Y_{T-\delta} \in L^2(\Omega, \F_{T-\delta}; \R^n)$ and $B, A_j \in 
Lip([T-\delta,T] \times \R^n; \R^n) \cap \mathcal{C}^{\infty}_b (\R^n; \R^n)$
for all $j = 1, \dots, d$.
We define
\[
e_p(\delta) := e^{ \delta^{p/2} ( \nB_1 + \nA_1 )^p },
\]
and
\[
N_k(A,B) = \nA_{k-1}^{k} \Bigl( \nB_k + \nA_k \Bigr)^{ (k+1)^2 }.
\]
The following proposition gives the conditional version of the estimates
in \cite{SDM}, Lemma 2.1 and Corollary 1.

\begin{prop}
For any $k \geq 1$ and any $p>1$ there exists a positive constant
$d_{k,p}$ depending on $k,p$ but not on the bounds on $B$ and $A$ and their derivatives such that, for every $T-\delta \le t \le T$ and for every
$l = 1,\dots,m$,
\be \label{e:boundDeriv}
\sup_{\alpha \in \Theta_k} \sup_{s_1,\dots,s_k \in [T-\delta, T)^k}
\mathbb{E} \Bigl[ \left| D^{k,\alpha}_{s_1,\dots,s_k} Y^l_t \right|^p \Big| \F_{T-\delta}
\Bigr]^{1/p}
\leq
d_{k,p} N_k(A,B) e_p(\delta)^{ d_{k,p} }
\ee
\be \label{e:boundTripleNorm}
||| Y^l_t |||_{T-\delta, \delta, k, p} \leq
2 k d^k d_{k,p} \times \sqrt{\delta} \times
N_k(A,B) e_p(\delta)^{ d_{k, p} }.
\ee
\end{prop}

\proof{
Inequality (\ref{e:boundDeriv}) relies on the same proof as Lemma 2.1 in \cite{SDM}.
Estimate (\ref{e:boundTripleNorm}) is a consequence of (\ref{e:boundDeriv}):
for any $i = 1,\dots,k$ and $l = 1,\dots,n$, we have
\[
\Bigl( \int_{[T-\delta,T)^i} |D^{i,\alpha}_{s_1, \dots, s_i} Y^l_t |^2 ds_1 \cdots ds_i \Bigr)^{p/2}
\le
\delta^{i(\frac p2-1)} \int_{[T-\delta,T)^i} |D^{i,\alpha}_{s_1, \dots, s_i} Y^l_t |^p ds_1 \cdots ds_i,
\]
hence, using (\ref{e:tripleNorm}) and $\sum_{i=1}^k card(\Theta_i) = \sum_{i=1}^k d^i \le k d^k$, we obtain
\[
\begin{aligned}
||| Y^l_t |||^p_{T-\delta, \delta, k, p} &\le
2^{p/2} (k d^k)^p \max_{i = 1, \dots, k} \delta^{i(\frac p2-1)} 
\max_{\alpha \in \Theta_i} \Bigl( \int_{[T-\delta,T)^i}
\esp[ |D^{i,\alpha}_{s_1, \dots, s_i} Y^l_t |^p | \F_{T-\delta} ] ds_1 \cdots ds_i \Bigr) 
\\
&\le
2^{p/2} (k d^k)^p \times \delta^{p/2} \times 
d_{k,p} N_k(A,B)^p e_p(\delta)^{ p d_{k,p} },
\end{aligned}
\]
which proves (\ref{e:boundTripleNorm}).
}
\endproof

\subsection{Proof of Proposition \ref{p:reminderEstim}}

\proof{
We assume without further mention that we are on the set $A_{\delta,l}$.
Since $\Td$ and $\delta$ are fixed, we drop them from the notation and write
$||F||_{k,p}$ instead of $||F||_{\Td,\delta,k,p}$ and so on.

We first show that estimate (\ref{e:reminderEstim}) holds for $\esp[  |R_{\Delta}|^p | \F_{T-\delta} ]^{1/p}$.
Using $|\eta(t,x) - \eta(s,y)|\sqrt{\psi(v)} \le K\sqrt y(|s-t|+|y-x|)$ and applying Burkholder's inequality we obtain
\[
\begin{aligned}
\esp[  |R|^p | \F_{T-\delta} ] &\le
c_p
\Bigl( \delta^{p-1} \int_{T-\delta}^T \esp[ | \psi(\Vbar_t) |^p | \F_{T-\delta} ] dt
\\
&\quad + \delta^{p/2-1}\int_{T-\delta}^T
\esp [ ( \eta(t, \Xbar_t)-\eta(T-\delta, X_{T-\delta}) )^p \psi(\Vbar_t)^{p/2} |
\F_{T-\delta} ] dt \Bigr)
\\
&\le c_p \Bigl( (\delta y)^{p}
+ \delta^{p/2-1} K^p y^{p/2} \int_{T-\delta}^T
( \delta^p + ||\Xbar_t - X_{T-\delta}||_{0,p}^p ) dt \Bigr)
\\
&\le c_p (\delta y)^{p/2}
\Bigl( (\delta y)^{p/2} + K^p \delta^p + K^p C_p \delta^{p/2} e^{C_p T^p}
\Bigr)
\\
&\le c_p (\delta y)^p e^{C_p T^p},
\end{aligned}
\]
where we have used $\delta < \sqrt{\delta} < \sqrt{\delta y}$ and Lemma
\ref{l:preliminary} to estimate
$||\Xbar_t - X_{T-\delta}||_{0,p}^p = \esp [ |\Xbar_t - X_{T-\delta}|^p | \F_{T-\delta} ]$.
Then we have
\be \label{e:zeroNorm}
|| \RD ||_{0,p} \le c_p \frac{1}{\rhop \etaUnder} \sqrt{\delta y} e^{C_p T^p}.
\ee
We now estimate the Sobolev norms of $\Xbar$ and $\Vbar$.
Notice that by the definition of $\psi$ we have
$|\psi |_0 \le y + \frac32 \le 2 y$ hence $| \psi (\cdot) |_k \le c^{(1)}_k y$
for all $k \ge 1$, for some constant $c^{(1)}_k$.
Similarly, $|\sqrt{\psi} |_0 \le \sqrt{y + \frac32} \le 2 \sqrt{y}$ and
$\frac{d}{dv} \sqrt{\psi(v)} = \frac{\psi'(v)}{2 \sqrt{\psi(v)}}$,
hence
\[
\Bigl| \frac{d}{dv} \sqrt{\psi(\cdot)} \Bigr|_0 \le \frac{ |\psi'|_0 }{ 2 \sqrt{y - \frac32} }
\le \frac{ |\psi'|_0 }{ \sqrt{2 y} }
\]
and it can be seen easily that $| \sqrt{\psi}(\cdot) |_k \le c^{(1)}_k \sqrt{y}$
for an eventually different constant $c^{(1)}_k$.
With a slight abuse of notation, we write $| \sigma \sqrt{\psi} |_k$ (resp.
$| \eta \psi |_k$) for the $|\cdot|_k$-norm of the function  $(t,v) \to \sigma(t, v) \sqrt{\psi(v)}$
(resp. $(t,x,v) \to \eta(t, x) \sqrt{\psi(v)}$).
Then, using (\ref{e:boundTripleNorm}) with $k=2$, for any $t \in [T-\delta,T]$ we have
\[
\begin{aligned}
||| \Vbar_t |||_{2,8p} &\le
4 \times d_{2,8p} \times \sqrt{\delta} \times
| \sigma \sqrt{\psi} |_1^2 ( |\beta|_2 + |\sigma \sqrt{\psi} |_2)^{9}
\\
& \quad \times
\exp ( d_{2,8p} \: \delta^{4p} ( |\beta|_1 + |\sigma \sqrt{\psi} |_1)^{8p} )  
\\
&\le
c^{(2)}_{p} K^{9} \times \sqrt{\delta} \times | \sqrt{\psi} |_2^{9} | \sqrt{\psi} |_1^{2} \times
\exp ( d_{2,8p} \: \delta^{4p} K^{16p} |\sqrt{\psi}|_1^{8p} )
\\
&\le
c^{(2)}_{p} \times \sqrt{\delta y^{11}} \times
\exp ( c^{(2)}_{p} \times (\delta y)^{4p} ).
\end{aligned}
\]
Hence, using $\delta y < 1$ we get $||| \Vbar_t |||_{2,8p} \le c^{(2)}_{p} \times
\sqrt{\delta y^{11}}$, for a (eventually different) constant $c^{(2)}_{p} $.
An analogous estimate holds for $||| \Xbar_t |||_{2,8p}$: observing that the only difference
is in the contribution of the drift term $|\eta \psi|_2 \le K c^{(1)}_3 y $, 
we have
\[
\begin{aligned}
||| \Xbar_t |||_{2,8p} &\le
c^{(2)}_{p} \sqrt{\delta} \times |\psi|_2^{9} \times |\sqrt{\psi}|_1^{2} \times
\exp ( c^{(2)}_{p} \times (\sqrt{\delta} |\psi|_1)^{8p} )
\\
&\le
c^{(2)}_{p} \sqrt{\delta y^{20}}
\end{aligned}
\]
since $\sqrt{\delta} y < 1$, too.
Now using (\ref{e:chain}) and denoting $\mu = \mu(2)$, we have
$|| \eta(t,\Xbar_t)^2||_{2,2p} \le \mu |\eta|_{2}^2 ||| \Xbar_t |||_{2,8p} $
and $|| \psi(\Vbar_t) ||_{2,2p} \le \mu |\psi|_{2} ||| \Vbar_t |||_{2,8p}
\le \mu c^{(1)}_3 y ||| \Vbar_t |||_{2,8p}$.
Hence, using (\ref{e:derivProd}), we get
\be \label{e:etaPsi}
\sup_{T-\delta\le t \le T} ||| \eta(t,\Xbar_t)^2 \psi(\Vbar_t)|||_{2,p} \le
2! 2^2 \mu^2 K^2 c^{(1)}_3 y
\times ||| \Xbar_t |||_{2,8p} ||| \Vbar_t |||_{2,8p} \le c_p \sqrt{\delta^2 y^{31}},
\ee
where the constant $c_p$ also depends on $K$.
We denote $R = -\frac12 I + J$, setting $I = \int_{T-\delta}^T \eta(t,\Xbar_t)^2 \psi(\Vbar_t) dt$ and $J = \int_{\Td}^T (\eta_t - \eta_{T-\delta}) \sqrt{\psi(\Vbar_t)} (\rho dW^1_t + \rhop dW^2_t)$.
Then, using (\ref{e:tripleNorm})
\[
\begin{aligned}
||| I |||^p_{2,p}
&\le
2^{p/2} \esp \Bigl[ \delta^{p-1} \int_{T-\delta}^T
\Bigl( 
\sum_{k = 1}^2 \sum_{\alpha \in \Theta_k}
\int_{[\Td,T)^k} | D_s^{k, \alpha} \eta(t,\Xbar_t)^2 \psi(\Vbar_t) |^2
\Bigr)^{p/2}
dt
\Big| F_{T-\delta} \Bigr]
\\
&\le
2^{p/2+1} \delta^{p-1} \int_{T-\delta}^T |||  \eta(t,\Xbar_t)^2 \psi(\Vbar_t) |||_{2,p}^{p} dt
\\
&\le
c_p (\delta^4 y^{31})^{p/2}.
\end{aligned}
\]
We now estimate the first Sobolev norm of $J$.
For the ease of notation, we write $\eta_t$ for $\eta(t,\Xbar_t)$.
For $\Td \le s \le t \le T$, we have $D^{1,j}_s((\eta_t - \eta_{T-\delta}) \sqrt{\psi(\Vbar_t)}) =
D^{1,j}_s(\eta_t \sqrt{\psi(\Vbar_t)})$, hence
\begin{multline}
D^{1,j}_s \int_{\Td}^T (\eta_t - \eta_{T-\delta}) \sqrt{\psi(\Vbar_t)} (\rho
dW^1_t + \rhop dW^2_t)
=
(\eta_s - \eta_{\Td}) \sqrt{\psi(\Vbar_s)}) \rho_j
\\
+ \int_{\Td}^T D^{1,j}_s(\eta_t \sqrt{\psi(\Vbar_t)}) (\rho
dW^1_t + \rhop dW^2_t)
\end{multline}
with $\rho_j = \rho 1_{j=1}+\rhop 1_{j=2}$.
Using bound (\ref{e:boundDeriv}) and proceeding as for (\ref{e:etaPsi})
we obtain \\ $\sup_{t \in [\Td,T]}\esp\Bigl[ \Bigl| D^{1,j}_s(\eta_t \sqrt{\psi(\Vbar_t)}) \Bigr|^p \Big| F_{\Td} \Bigr]^{1/p} \le c_p |\eta|_1 |\sqrt{\psi}|_1 \sqrt{y^{11+20}} \le
c_p \sqrt{y^{32}}$, where $c_p$ depends also on $K$.
Then, starting from (\ref{e:tripleNorm}) and using Burkholder's
inequalities
\[
\begin{aligned}
||| J |||_{1,p}^p
&\le
2^{p} \delta^{p/2-1} 
\int_{\Td}^T || ( \eta_s - \eta_{T-\delta}) \sqrt{\psi(\Vbar_s)} ||_{0,p}^p ds
\\
&\quad+
\delta^{p-1} \sum_{j =1,2}
\int_{T-\delta}^T \int_{T-\delta}^T \esp\Bigl[ | D^{1,j}_s(\eta_t \sqrt{\psi(\Vbar_t)}) |^p \Big| F_{\Td} \Bigr] dt ds
\\
&\le
2^{p} K^p \delta^{p/2-1}
y^{p/2}
\int_{\Td}^T (\delta + || \Xbar_s - X_{\Td} ||_{0,p})^p ds
+
c_p \delta^{p+1} y^{32p/2}
\\
&\le
c_p ( \delta^2 y^{32})^{p/2} e^{C_p T^p}. 
\end{aligned}
\]
The Sobolev norms of higher order are estimated in a similar way,
giving the bound $||| J |||_{2,p} \le c_p \sqrt{ \delta^2 y^{32}} e^{C_p T^p}$.
Finally,
$||| R |||_{2,p} \le ||| I |||_{2,p} + ||| J |||_{2,p} \le c_p \sqrt{ \delta^2 y^{32}} e^{C_p T^p}$ and this last estimate together with (\ref{e:zeroNorm}) yields
$||| R_{\Delta} |||_{2,p} \le c_p \frac1{\rhop \etaUnder} \sqrt{ \delta y^{31}} e^{C_p T^p}$,
which is (\ref{e:reminderEstim}).
}
\endproof

\bibliographystyle{plain}
\bibliography{References}

\end{document}